\documentclass[12pt]{article}
\pdfoutput=1
\usepackage[nosort]{cite}
\usepackage{pifont}

\usepackage{times}
\usepackage{tikz}

\usepackage{comment}
\usepackage{booktabs}
\usepackage[utf8]{inputenc}
\usepackage{amsmath}
\usepackage{amssymb}
\usepackage{mathtools}
\numberwithin{equation}{section}
\usepackage{slashed}
\usepackage{braket}
\usepackage[svgnames]{xcolor}
\usepackage{url}
\usepackage[colorlinks,citecolor=DarkGreen,linkcolor=FireBrick,linktocpage]{hyperref}
\usepackage{cite}
\usepackage{graphicx}
\usepackage{float}
\usepackage{subcaption} 
\usepackage{courier}
\usepackage{bm}

\usepackage{dashbox}
\usepackage{caption}
\usepackage{subcaption}
\usepackage{enumitem}
 
\usepackage{footmisc}
\usepackage{mdframed}

\usepackage[margin=1in]{geometry}
\usepackage{physics}
\usepackage{hyperref}
\usepackage{microtype}

\usepackage{blkarray}
\usepackage{arydshln}
\usepackage{dsfont}

\usepackage{calc}

\usepackage{accents}

\newcommand{\ie}{\begin{equation}\begin{aligned}}
\newcommand{\fe}{\end{aligned}\end{equation}}

\renewcommand{\title}[1]{\vbox{\center\LARGE{#1}}\vspace{5mm}}
\renewcommand{\author}[1]{\vbox{\center#1}\vspace{5mm}}
\newcommand{\address}[1]{\vbox{\center\em#1}}

\makeatletter
\newsavebox{\@brx}
\newcommand{\llangle}[1][]{\savebox{\@brx}{\(\m@th{#1\langle}\)}%
  \mathopen{\copy\@brx\kern-0.5\wd\@brx\usebox{\@brx}}}
\newcommand{\rrangle}[1][]{\savebox{\@brx}{\(\m@th{#1\rangle}\)}%
  \mathclose{\copy\@brx\kern-0.5\wd\@brx\usebox{\@brx}}}
\makeatother

\begin{document}

\begin{titlepage}

    \title{Trapping $\tfrac{h}{2e}$ Flux in Metals }

    \author{Zohar Komargodski, Fedor K. Popov}

    \address{Simons Center for Geometry and Physics, Stony Brook University, Stony Brook, NY}
    \abstract
    We report on a new flux quantization phenomenon in metals.
    We study the response of normal metals to the presence of localized magnetic flux.
    We find that, due to backreaction effects, the metal traps 0 flux or $\tfrac{h}{2e}$ flux (half flux). We exhibit this effect both for metals pierced by magnetic solenoids and metals wrapping a magnetic solenoid. In the latter case we  demonstrate the trapping of magnetic flux analytically.
    Furthermore, we find that as the solenoid is adiabatically turned off, a logarithmically enhanced localized equilibrium current persists, reflecting perfect defect-diamagnetism of the Fermi gas.
\end{titlepage}

\eject

\setcounter{tocdepth}{3}
\tableofcontents

\section{Introduction and Summary}

\begin{figure}[htbp]
    \centering
    \includegraphics[width=0.45\textwidth]{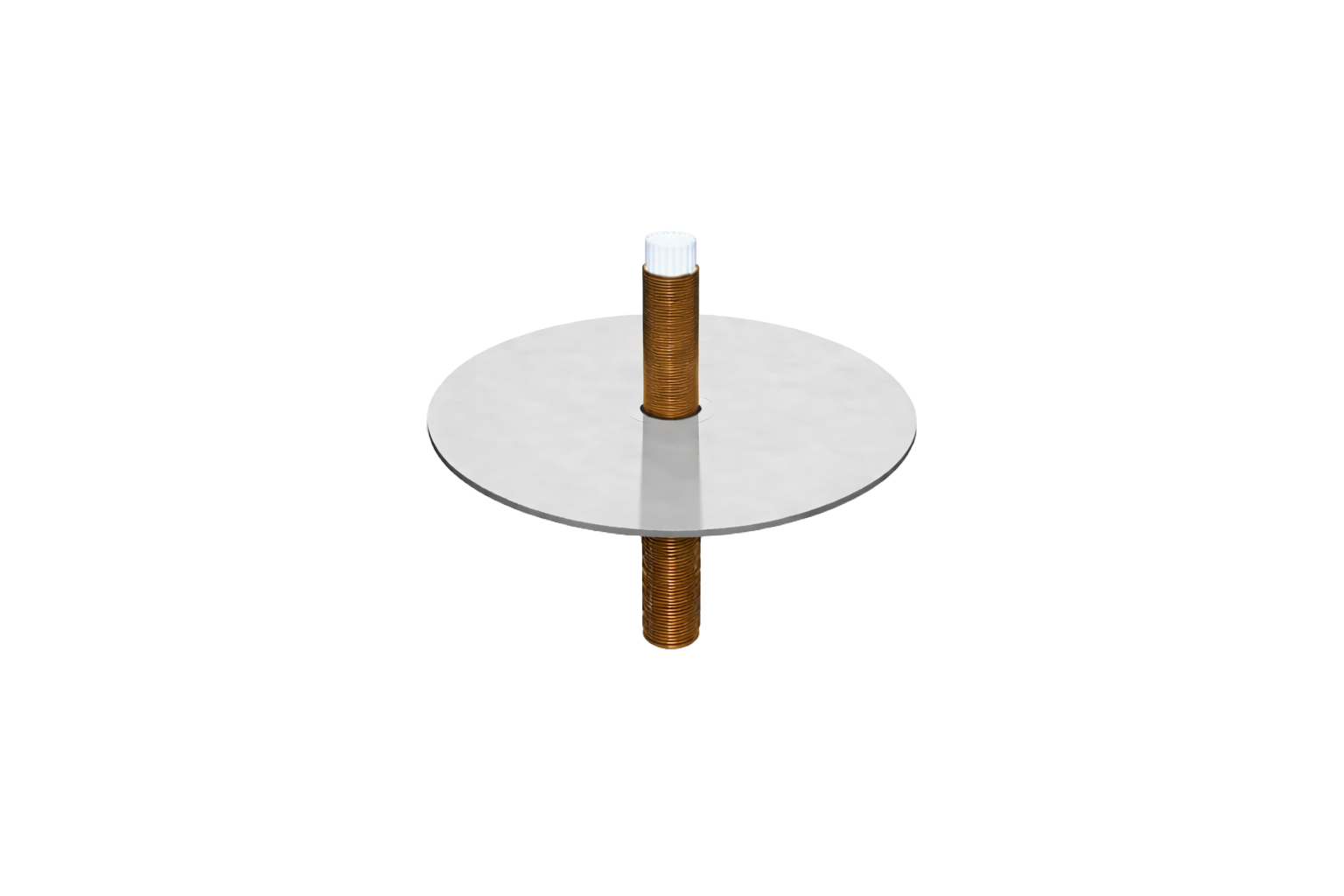}
    \includegraphics[width=0.45\textwidth]{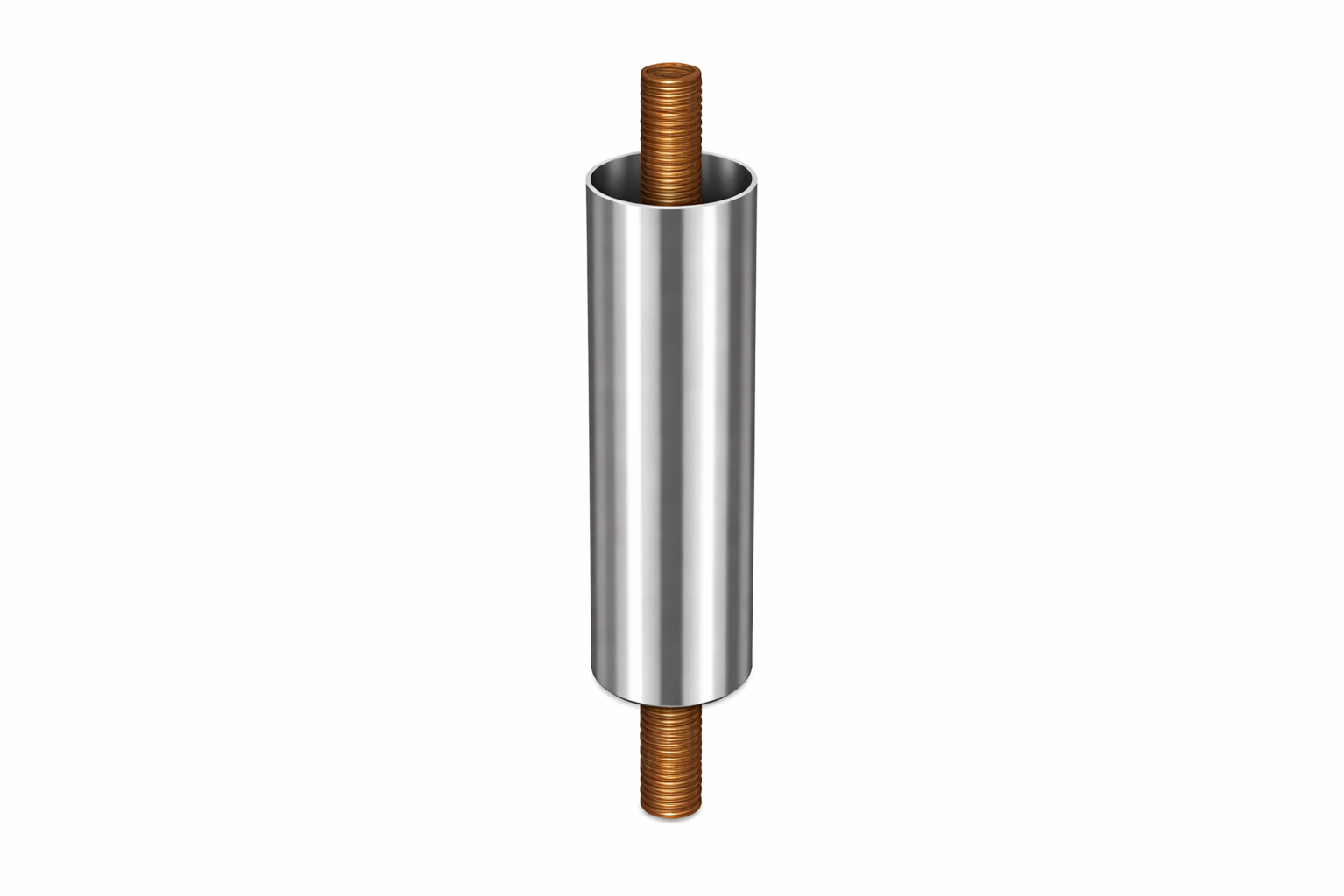}
    \caption{Two solenoid configurations. Left: A magnetic solenoid with flux $\Phi$ piercing a metal. Right: A magnetic solenoid with flux $\Phi$ wrapped by a metal.}
    \label{fig:solenoids}
\end{figure}

The problem of defects in gapless systems attracted much attention recently. Here we consider the fate of the Aharonov-Bohm solenoid piercing a normal metal. We will also discuss a cylindrically shaped metal wrapping the solenoid. The two setups are depicted in Figure~\ref{fig:solenoids}.
In a metal, particles can effectively appear and disappear from the Fermi sea so the situation is more involved than the textbook problem of a single particle around an Aharonov-Bohm solenoid~\cite{PhysRev.115.485,PhysRev.123.1511} (see also~\cite{batelaan2009aharonov} for a review).

The solenoid leads to no electromagnetic fields outside of it, nevertheless particles are excited from the Fermi sea and a current flows in the system around the solenoid. This happens as well in quantum field theories (where particles are excited from the relativistic vacuum)~\cite{Soderberg:2017oaa,Giombi:2021uae,Bianchi:2021snj,Gimenez-Grau:2021wiv, Dowker:2022mex, SoderbergRousu:2023pbe,Barkeshli:2025cjs}.
In metal rings, these vacuum currents are known as ``persistent currents''~\cite{buttiker1985generalized,cheung1988persistent,imry2002introduction} and they have been observed in mesoscopic systems~\cite{PhysRevLett.102.136802,doi:10.1126/science.1178139}.

Unlike the familiar setup of a thin metal ring, here we consider the geometries of Figure~\ref{fig:solenoids}. We will see that in both of these setups a qualitatively new phenomenon, of {\it flux quantization}, arises.
The flux quantization that we discover in these geometries is somewhat reminiscent of superconductors, but it is very different in the details.  The currents swirling in the metal create their own induced magnetic fields $\vec B_{\rm ind}$ with vector potential $\vec A_{\rm ind}$ that in turn affects the electrons wave functions and leads to a strong backreaction effect\footnote{In this paper we treat the Maxwell field classically while the fermions are treated quantum mechanically. We also note that in the problem of the Condon domains~\cite{gordon2003magnetic} a similar self-consistent calculation is done. We thank Leonid Levitov and Subir Sachdev for discussion and these comments.}.

Common lore suggests that this backreaction is negligible since the induced magnetic fields are suppressed by the fine structure constant and the Fermi velocity. However, in the setup of the solenoid wrapped by a metal of Figure
~\ref{fig:solenoids}, it turns out that the system develops a new length scale
\begin{gather}\label{trapscale}
    \lambda_{\rm trapping} \sim \frac{R^\frac14}{\sqrt{\alpha k_F^\frac32 \frac{v_F} {c}}}~,
\end{gather}
with $\alpha$ the fine-structure constant, and $R$ the radius of the cylinder. The emergence of this length scale is very similar to electric charge screening in metals, though here it pertains to the magnetic flux. This length scale  governs the trapping of half flux or zero flux.
In other words, not only the backreaction is non-negligible, it leads to flux quantization. Whether the system prefers to trap
$h/2e$ flux or zero flux depends on the value of $k_FR$.
Similarly, we find very strong effects from the induced magnetic fields in the left geometry of Figure~\ref{fig:solenoids}, leading to the same flux quantization but with a different length scale.

\begin{figure}
    \centering
    \includegraphics[scale=0.6]{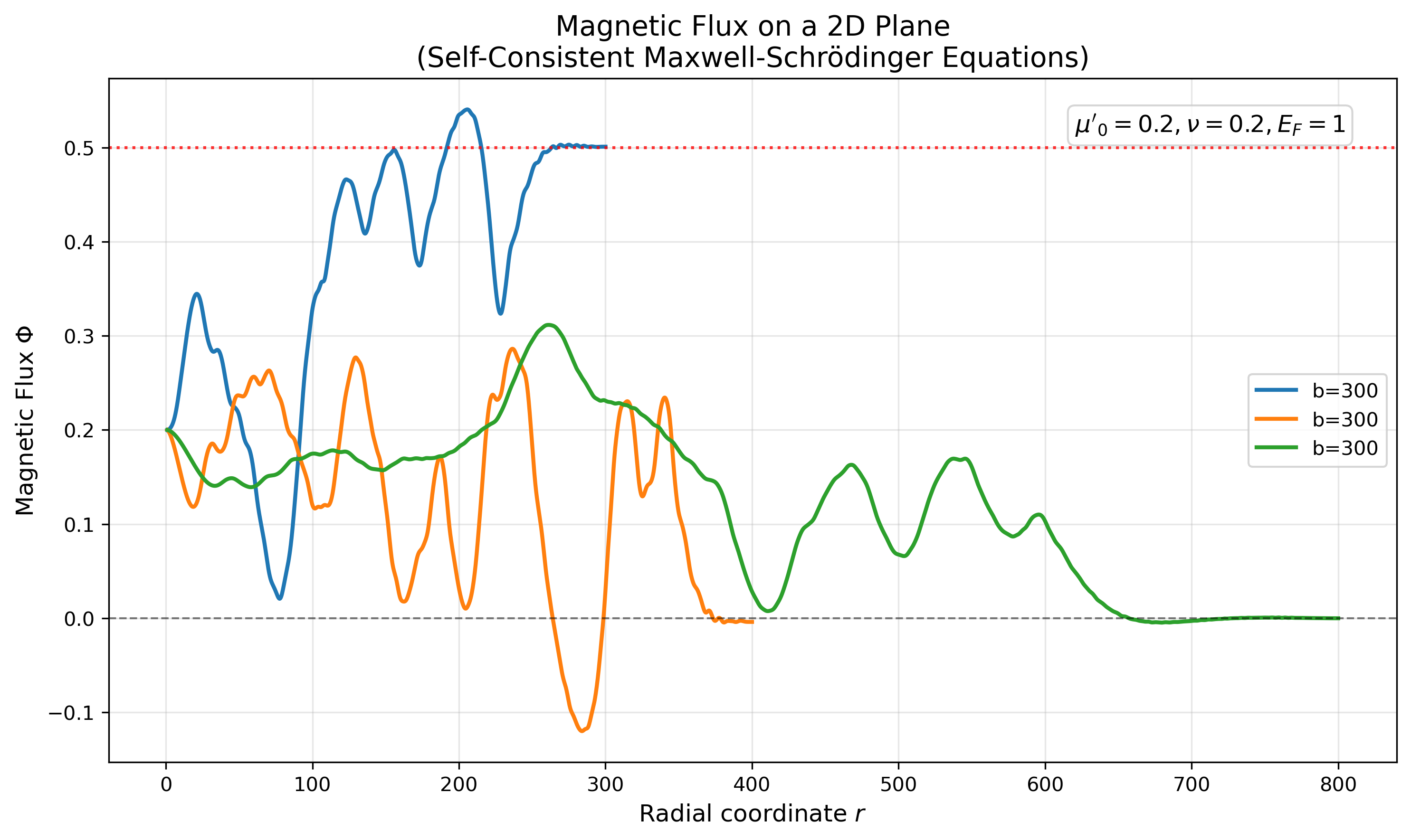}
    \caption{Examples of the self-consistent solution of Schr\"odinger and Maxwell equations. Depending on the size of the disk the flux could screen to $\Phi=0$ or to $\Phi/\Phi_0=\frac12$. }
    \label{fig:curscreen}
\end{figure}

\begin{figure}
    \centering
    \includegraphics[scale = 0.5]{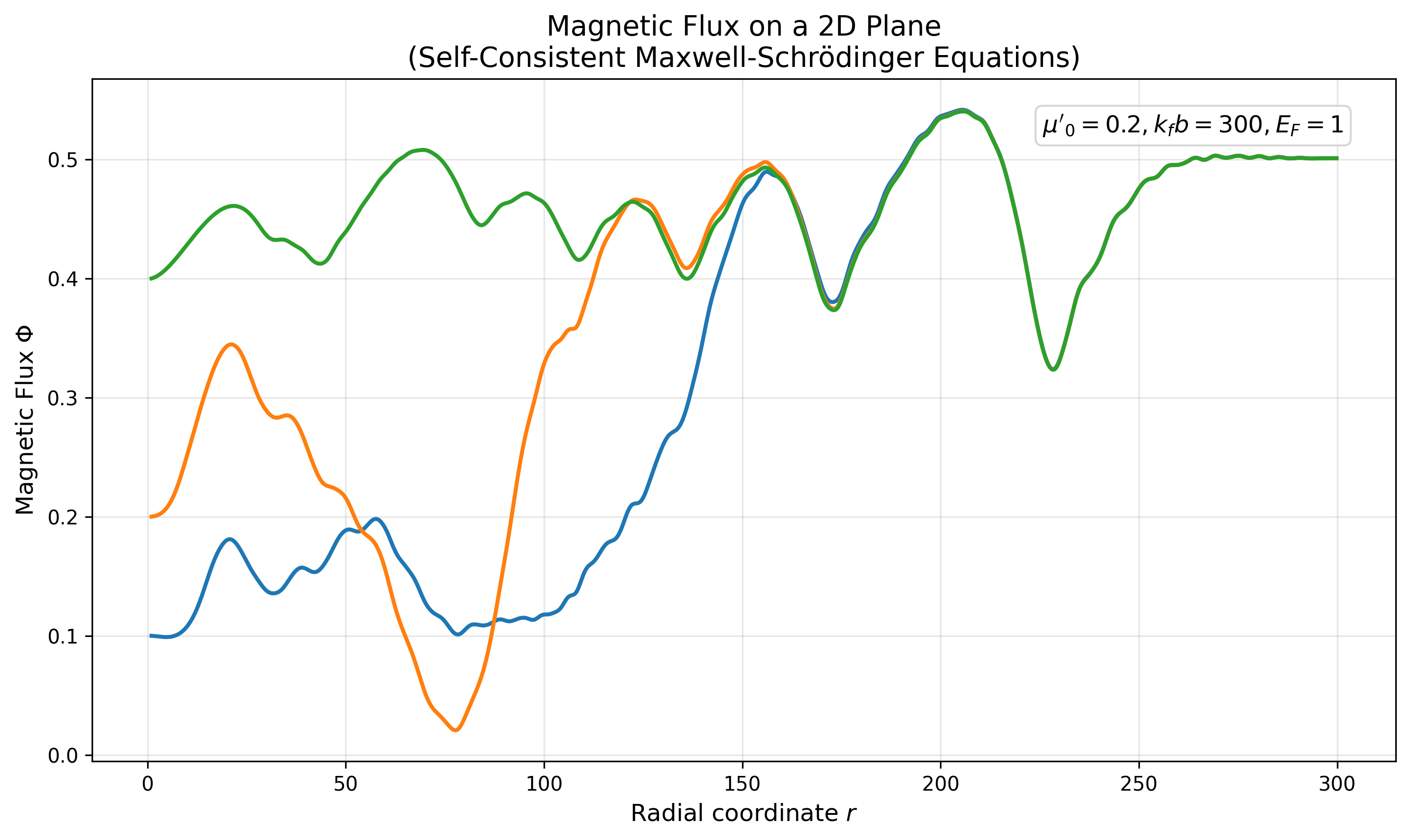}
    \caption{Examples of the self-consistent solution of Schr\"odinger and Maxwell equations. We see that near the boundary the fluxes do not depend on the initial value of the holonomy. At large $r$ the three lines are on top of each other and all correspond to trapping of half flux.}
    \label{fig:curscreenfixedb}
\end{figure}

In the cylinder geometry, we can show the emergence of the trapping scale~\eqref{trapscale} analytically.
By contrast, in the disk we do not have analytic results and demonstrate the trapping of half or zero flux numerically only.  In Figure~\ref{fig:curscreen} we show $\Phi$ is screened to 0 or half-flux depending on the system's size, $b$.
In Figure~\ref{fig:curscreenfixedb} we show how for a specific (fixed) system size, the flux is screened to half-flux independently of the original solenoid flux. We give a speculative explanation for the phenomenon of flux quantization, related to certain finite size effects on the disk.

In summary, a metal disk can serve effectively as a $h/2e$ flux solenoid.
Famously,  superconductors can trap $h/2e$ units of flux (the Abrikosov flux tube). It appears that ordinary metals may behave analogously, though the details are very different.
That the magnetic flux in metals is quantized is very surprising.

Another question that we discuss in this manuscript concerns with the limit that the solenoid becomes infinitely narrow. Thus, we will  take $k_F a\to 0$ and we will discuss what happens as the solenoid is adiabatically turned off, i.e.
$\Phi \ll \Phi_0$.
We find that the ground state energy is non-analytic.
The non-analyticity represents a break-down of linear response theory\footnote{This has to be contrasted with Landau's diamagnetism where the ground-state energy is analytic with respect to the overall applied magnetic field.}.
A nonzero current survives as $\Phi\to0$. That this current does not go away as we remove the solenoid is interpreted as perfect defect-diamagnetism, since the energy cost of external flux is proportional to the absolute value of the external flux. Therefore, there is a strong expulsion force in the Fermi gas against solenoids.\footnote{Similarly, the energy of a superconductor is proportional to the absolute value of the magnetic flux for small enough external magnetic flux.}

The current which survives the $\Phi\to0$ limit is localized mostly in the region around the solenoid
\begin{equation}\label{introexp}
    r \lesssim k_F^{-1} e^{-\frac{\Phi_0}{\Phi}}~.
\end{equation}
This short distance scale  is outside of the effective theory of the Fermi surface. Therefore, the localized current is non-universal. However, in metals with a small Fermi surface such as semi-metals it might be possible to observe this anomalously large current. At finite solenoid size $k_Fa$, the phase transition turns into a crossover.
Thus, we obtain {\it log-enhanced diamagnetism}. In practice, it means that
for small solenoid flux
\begin{equation}
    {\Phi\over \Phi_0}\approx {-1\over \log(k_F a)}\ll1
\end{equation}
we observe a very large current, of the order of $E_{F}/\Phi_0$, concentrated on the inner edge of the metal.
Figure~\ref{fig:sum_eigenvalues}  displays the energy of the ground state as a function of $\Phi$ for some values of  inner radius $a$ and outer radius $b$. We see how the result approaches the non-differentiable free energy at $k_F a\to0$ as we make the solenoid smaller and smaller.

The outline of the paper is as follows. In Section~2 we fix conventions, units, and some details of the setups to be studied. In Section~3 we analyze the persistent current response of an annular metal pierced by an Aharonov-Bohm solenoid while neglecting the backreaction. We emphasize the oscillatory current profiles and strong finite-size effects. In Section~4 we incorporate the induced magnetic fields and solve the coupled Schr\"odinger--Maxwell problem self-consistently: we first demonstrate analytically in the cylindrical geometry that the flux is attracted either to zero or to half-flux and identify the associated trapping length scale, and we then show numerically that on a finite disk the total flux is likewise driven to $0$ or to $1/2$. In Section~5 we study the limit of a thin solenoid and small imposed flux, where the ground-state energy becomes non-analytic and a log-enhanced equilibrium current persists as the solenoid is adiabatically removed, which we interpret as defect-localized perfect diamagnetism.

Various technical details are deferred to the appendices, including the analysis of boundary conditions at the solenoid, RPA calculations for electric and magnetic response, and derivations of the current formulas used in the main text.

\section{Setup and Conventions}
\label{app:notations}
Let us briefly set up the conventions that we will use throughout the paper.
Since we will study only stationary configurations, to make the choice of units transparent, we begin with the energy functional
\begin{gather}
    E[\hat{\psi}, A] = \int d^3 x  \left\langle\frac{\hbar^2 }{2m} \left|\left(\nabla - \frac{i e}{\hbar c} A\right) \hat\psi \right|^2 - E_F \left|\hat \psi\right|^2 \right\rangle + \int d^3 x \frac{\left(\nabla \times A\right)^2}{8\pi} ~.
\end{gather}
It will be useful to switch to dimensionless coordinates where everything is measured in terms of Fermi momentum $\hbar k_F$. Thus, we define
\begin{gather}
    x=k_F^{-1}x'~,\quad \hat \psi=k_F^{3/2} \hat \psi'~,\quad A={k_F\hbar c\over e}A'~, \quad j = \frac{e\hbar k_F^4}{m} j'~, \quad  E  = \frac{\hbar^2 k_F^2}{2m} E'.
\end{gather}
Now, $x',\hat \psi',A'$ are all dimensionless. Importantly, the expression of the particle number charge is unchanged and remains $\int d^3x  \bigl|\hat \psi\bigr|^2 =\int d^3x'  \bigl|\hat \psi'\bigr|^2$.
The energy functional becomes
\begin{equation}
    E'[\hat{\psi}', A'] =\int d^3x' \left\langle
    \bigl|( \nabla' - i  A')\hat \psi'\bigr|^2 - \bigl|\hat \psi'\bigr|^2
    \right\rangle+ \int d^3 x'\frac{(\nabla' \times A')^2}{4\pi \alpha {v_F\over c}}\,
    .
\end{equation}
In these conventions we obtain
\begin{equation}
    \nabla'\times B'=\mu'_0 \langle \hat J' \rangle~,\quad  \mu'_0 = 4\pi \alpha  \frac{v_F}{c}
\end{equation}
with
$$\hat{J}' = {i\over 2}
    \left(
    \hat{\psi}'^{*}\nabla'\hat{\psi}'
    - \hat{\psi}'\nabla'\hat{\psi}'^{*}
    - 2i A'\,|\hat \psi'|^2
    \right)~.
$$
Above $\alpha={e^2\over \hbar c}$, which is the usual fine structure constant.
In these dimensionless conventions it is manifest that the induced magnetic fields are suppressed by  $\alpha{v_F\over c}$ and are therefore very small. Henceforth we use these dimensionless quantities without the ' superscript.
In these conventions the holonomy of the gauge field is defined modulo the integers. We denote the fractional part by $\nu$ throughout the paper:
\begin{equation}{1\over 2\pi}\oint A = \nu \ {\rm mod}\  \mathbb{Z}~.\end{equation}
Equivalently,  $\nu= \Phi/\Phi_0$ modulo the integers with $\Phi_0=h/e$ the usual Aharonov-Bohm quantum.

We will sometimes refer to a ``2D setup,'' which is often technically simpler. By 2D we mean a thin-film sample of thickness $\delta$ whose lateral dimensions $L$ are much larger, $L\gg \delta$. In this limit the electronic degrees of freedom are effectively confined to the plane. To suppress the tangential magnetic fields, we assume that the sample is sandwiched between two
media with very large magnetic permeability, $\mu\gg 1$.
In this configuration the magnetic induction is effectively guided normal to the film,
so we may neglect the in-plane components and keep only the perpendicular one
$\mathbf{B} \approx B_{\perp}\,\hat{\mathbf{n}}$, where $\hat{\mathbf{n}}$ is the unit vector normal to the film. This follows from Maxwell’s boundary conditions at an interface between two media of different magnetic permeabilities $B_\perp = B'_\perp, H_\parallel = H'_\parallel$.
Since in a high-permeability medium we have $H_\parallel = B_\parallel/\mu \approx 0$, any tangential component of the magnetic field on the boundary must vanish. Thus only the perpendicular field component survives. We denote  $B_\perp \equiv B$, and we will treat it just like a scalar, satisfying
\begin{gather}\label{Ampere}
    \epsilon_{ab} \partial_b B = \mu_0' j_a,
\end{gather}
which is essentially the $2+1$ dimensional Maxwell equation.

\section{Current Oscillations on a Disk }\label{nobck}
Let us see how persistent currents behave in an annulus-shaped 2D metal bounded by the circles $r= a$ and $r= b$ with flux $\nu$ inserted in the middle (see the left panel of Figure~\ref{fig:solenoids}). In this section we ignore the backreaction of the induced magnetic fields. We solve the Schr\"odinger equation with Dirichlet boundary conditions and a solenoid at the center
\begin{gather}
    \left(- \frac{1}{r} \frac{\partial}{\partial r}\left( r \frac{\partial}{\partial r}\right) - \frac{\partial_\theta^2}{r^2} \right)\psi(r,\theta,z) = E \psi(r,\theta,z),
    \quad \psi(r,\theta + 2\pi,z) = e^{2 \pi i\nu}\psi(r,\theta,z)~,\notag\\
    \psi(a,\theta,z) = \psi(b,\theta,z) = 0~.
\end{gather}
The solution is a linear combination of the following form
\begin{gather}
    \psi(r,\theta,z)  = N_{n,p} \left( J_{n+\nu}(k_{n,p} r) Y_{n+\nu}(k_{n,p} a) -  J_{n+\nu}(k_{n,p} a) Y_{n+\nu}(k_{n,p} r)\right) e^{i (n + \nu) \theta }
\end{gather}
And the spectrum $k_{n,p}$ is obtained from
\begin{gather}
    J_{n+\nu}(k_{n,p} a)Y_{n+\nu}(k_{n,p} b) =  J_{n+\nu}(k_{n,p} b)Y_{n+\nu}(k_{n,p} a)
\end{gather}
Normalizing the wave functions \begin{gather}
    N_{n,p}^{-2} = \int\limits^b_a 2\pi r dr \left(J_{n+\nu}(k_{n,p} r) Y_{n+\nu}(k_{n,p} a) -  Y_{n+\nu}(k_{n,p} r)  J_{n+\nu}(k_{n,p} a)\right)^2
\end{gather}
we can readily calculate the azimuthal current density\footnote{
    Everywhere in this note we will be using a convention where $j_\theta = \frac{\hbar^2}{m} \operatorname{Im}\left[\psi^* \partial_\theta \psi\right]$ and $j^\theta = \frac{1}{r^2}j_\theta$.
}
\begin{gather*}\label{currentnofd}
    j^\theta(r) = \frac{1}{\pi r^2}\sum_{k_{n,p} \leq 1}  (n+\nu) N_{n,p}^2 \left(J_{n+\nu}(k_{n,p} r) Y_{n+\nu}(k_{n,p} a) -  Y_{n+\nu}(k_{n,p} r)  J_{n+\nu}(k_{n,p} a)\right)^2 %
\end{gather*}

It is possible to obtain an analytic result in 2D for the current in the infinite volume limit with a thin inner radius, i.e. $b\to \infty, a\to 0$.  We can define this limit by setting Dirichlet boundary conditions at $r=a$ and then send $a\to 0$. One can also approach it  as a problem of looking for fixed points of the non-relativistic defect renormalization group.
This is presented in detail in appendix~\ref{bc}.
The bottom line of that analysis is that the
modes with angular momentum $\ell\in \mathbb{Z}+\nu$ have boundary condition $\Psi \sim r^{|\ell|}$ near the origin. This is a fixed point of the renormalization group of the electron wave functions, preserving the Schr\"odinger symmetry algebra \cite{Kaplan:2009kr,Boisvert:2025hex,Raviv-Moshe:2024yzt}.

The simplification from taking $a\to0$ is that only the $J$ Bessel functions survive at the infrared stable fixed point of the defect renormalization group (other boundary conditions exist too, as we explain in appendix~\ref{bc}).
If we also take the $b\to\infty$ limit, i.e. that the metal is very large, the normalization of the wave functions dramatically simplifies.
Indeed, we obtain the wave functions\footnote{The normalization of the wave functions is such that $$ \int rdrd\theta \Psi_{\ell,E}\Psi_{\ell',E'} = {\pi\over b}\delta_{\ell \ell '} \delta(k-k')~.$$
At finite volume ${\pi\over b}\delta(0)$ should be interpreted as 1, as required. }
\begin{gather}
    \Psi_{\ell,E} = \sqrt{\frac{k}{2b }} e^{i \ell\theta} J_{|\ell|}(kr)
\end{gather}
and the energy is given by $E=k^2$.
These wave functions obey the required boundary conditions.

Finally, we have to determine the density of states. At large volume the density of states is  $\ell$ independent and is given by
${dN_{\ell}\over dk}={b\over \pi}$.
Therefore in the infinite volume limit and $a\to 0$ we find the following expression for the current
in 2D\footnote{The derivation is presented in appendix~\ref{app:cur_AB_2d}}
\begin{gather}
    j^\theta(r) = \frac{1}{r^2}\sum_{\ell \in \mathbb{Z} + \nu} \ell \int\limits^{1}_0 \frac{dk}{2\pi} k  J^2_{|\ell|}(k r) =   \notag\\
    =\frac{1}{24 \pi r^3}\Bigg[
        r\left(1-2r^{2}-3\nu+2\nu^{2}\right)J_{1-\nu}(r)^{2}
        + 2(-1+2\nu)\left(-r^{2}+(-1+\nu)\nu\right)J_{1-\nu}(r)\,J_{-\nu}(r) - \notag\\
        -r\left(2r^{2}+\nu-2\nu^{2}\right)J_{-\nu}(r)^{2}
        +r\left(2r^{2}+(5-2\nu)\nu\right)J_{\nu}(r)^{2}
        + \\
        +2\left(\nu+\nu^{2}(-3+2\nu)-x^{2}(1+2\nu)\right)J_{\nu}(r)\,J_{1+\nu}(x)
        +r\left(-1+2r^{2}+(3-2\nu)\nu\right)J_{1+\nu}(r)^{2}
        \Bigg]~. \notag
\end{gather}
At large distances we get that the current behaves as
\begin{gather}\label{2DAs}
    j^\theta(r)  \propto \frac{(1 - 2\nu) \sin(\pi \nu) \sin(2 r) }{8 \pi r^4}~.
\end{gather}
Importantly, the current identically vanishes for $\nu=0,1/2 \ {\rm mod}\ 1$. At $\nu=0$ it vanishes because the metal is in the usual ground state while at $\nu=1/2$ it vanishes by  parity (or charge conjugation symmetry).
We observe that the current oscillates with  period $\lambda_F =\pi$ and decays in magnitude.
These oscillations are analogous to Friedel oscillations of charge density.

\begin{figure}[t]
    \centering
    \includegraphics[scale = 0.6]{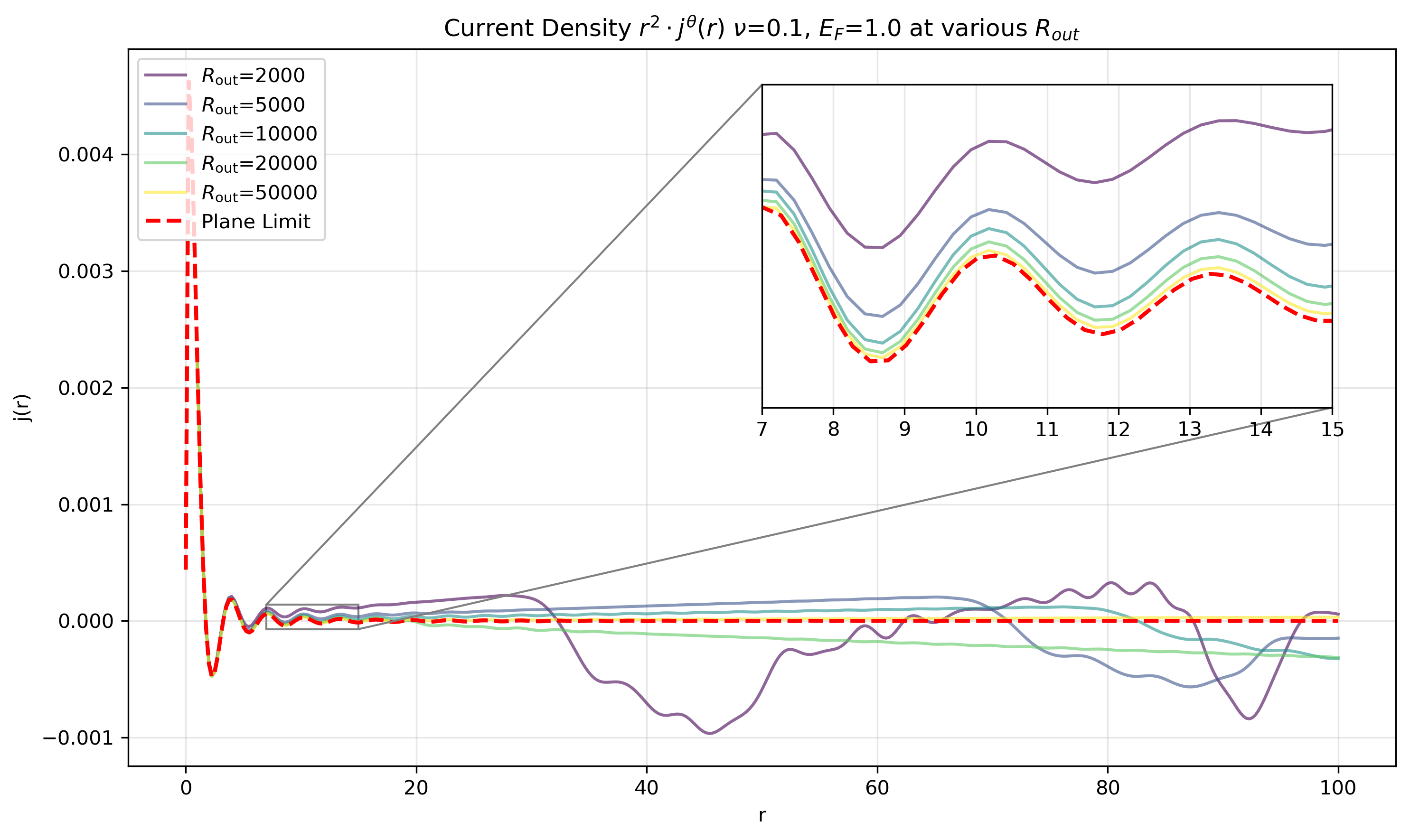}
    \caption{Radial dependence of the scaled persistent current density $r^2 j^\theta(r)$ for a 2D Fermi gas with AB flux $\nu = 0.1$ and Fermi energy $E_F = 1.0$. The red dashed line represents the analytical thermodynamic limit~\eqref{2DAs} (infinite plane, $b=R_{\mathrm{out}} \to \infty$). Solid lines display numerical results for varying finite system sizes ($R_{\mathrm{out}}$). As $R_{\mathrm{out}}$ increases, the finite-size effects diminish, and the numerical solution converges toward the theoretical thermodynamic limit prediction.}
    \label{fig:currents}
\end{figure}

A remarkable fact is that the large volume limit is reached extremely slowly as we send the outer boundary to infinity $b\to\infty$. In Figure~\ref{fig:currents} we plot examples of the results.
The different colors in Figure~\ref{fig:currents} correspond to different values of $b$.
We see that even for enormous systems, with size, say, $10^4$ in Fermi units, the infinite volume limit is only visible around $r=10$. Beyond that we see a linear term, which is a finite size effect, that gives a contribution of the order $r^2j_{\rm out}^\theta \sim \frac{r}{b }$ and  becomes large in comparison to the infinite disk limit where $r^2j^\theta \sim \frac{1}{r^2}$. Hence at distances $r \sim b^\frac13\ll b$ we see very large finite size effects.  These finite size effects are presumably responsible for the surprising quantization of flux that we will see in the next section.

Something quite interesting also happens on distance scales  $r\ll k_F^{-1}$, that we discuss later. As we decrease $\nu$, the spike of the current becomes more narrow but also bigger near the origin, suggesting in the limit $\nu \to 0$ the current becomes a delta function, that leads to the non-analyticity of the ground state energy, as we will discuss in section~\ref{secanomaly}.

\section{Self-Consistent Oscillations and Flux Quantization}\label{secselfconsistent}

\subsection{General Considerations}\label{RPAreview}
In the previous section we have considered the oscillating currents caused by a solenoid. However, we did not take into account the induced magnetic fields that arise due  Amp\'ere's law~\eqref{Ampere}. These induced magnetic fields  affect the electron dynamics and a self-consistent calculation is required.

We now briefly review the electron backreaction effects in ordinary, infinite 3D metals. We integrate out the electrons and study the effective theory of the magnetic field.
We expand the effective potential for the magnetic field to second order in the gauge field. This approach coincides with the Random Phase Approximation (RPA).

Ignoring the electrons in the metal, we need to minimize the usual energy (in Coulomb gauge)
\begin{align}
    \mathcal E[\mathbf A]
     & =
    \int {d^3q\over (2\pi)^3}\left[ -\frac{1}{c}\mathbf J_{\rm ext}(\mathbf r)\cdot\mathbf A(\mathbf r)
    +{1\over 8\pi} q^2|\bm A(\bm q)|^2\right]~,
\end{align}
the presence of the metal modifies (in the RPA approximation) the energy to
\begin{align}\label{RPAquote}
    \mathcal E[\mathbf A]
     & =
    \int {d^3q\over (2\pi)^3}\left[ -\frac{1}{c}\mathbf J_{\rm ext}(\mathbf r)\cdot\mathbf A(\mathbf r)
    +{1\over 8\pi} q^2|\bm A(\bm q)|^2\left(1+{ne^2\over mc^2}{f_T(z)\over q^2}\right)\right]~.
\end{align}
where $z=q/2k_F$ and
\begin{align}
    f_T(z)=\frac{3}{8}\qty(1+z^2)-\frac{3}{16z}\,(1-z^2)^2\ln\abs{\frac{1+z}{1-z}}.
    \label{eq:KTstaticmain}
\end{align}
The derivation is reviewed in detail in Appendix~\ref{RPAappendix}.

Since for $z\ll 1$, $f_T(z)=\frac{q^2}{4k_F^2}-\frac{q^4}{80k_F^4}+ \cdots$,
what the electrons in the metal do to the long distance physics is encompassed in an effective permeability.
For instance, for constant magnetic fields, the metal
increases the energy cost to establish the field  by $\Delta \mathcal E[\mathbf A]_{matter} = - \chi_{\mathrm{Landau}} B^2$ and we can read out the standard Landau diamagnetic susceptibility as
\begin{align}
    \chi_{\mathrm{Landau}}=-\frac{e^2 k_F}{24\pi^2 m c^2}~.
\end{align}

For a solenoid, as we considered in the previous section, the singularity of $f_T(z)$ at $z=1$ is responsible for the oscillations we have seen, but within the RPA nothing much happens beyond that, and in particular, the gauge field holonomy does not decay at long distances.
Indeed a constant holonomy configuration is a solution of~\eqref{RPAquote}. To see that, plug in a localized source, corresponding to a current independent of the $z$ direction and localized in the $x,y$ plane
$\mathbf{\bm J}_{\rm ext}(\bm q)\sim \nu (q_y\delta(q_z),-q_x\delta(q_z),0)$. Solving for the gauge field from~\eqref{RPAquote} we obtain a gauge field with a nonzero asymptotic holonomy.

This must be contrasted with the situation for electric sources, also reviewed in appendix~\ref{RPAappendix}, where the analog of $f_T(z)$ has a constant term in its  expansion around $z=0$ and leads to screening of electric fields.

In summary, what the RPA teaches us is that the holonomy due to solenoids is {\it not screened} in metals in infinite volume in flat space\footnote{By contrast, in materials which are not Fermi gases where $f_{T}(z)$ has a power smaller than $z^2$ in its expansion around the origin, the holonomy would be screened.}.
This is not the end of the story, though. First of all, as we will shortly see, in cylindrical geometry, in contrast to the plane, the holonomy is screened due to a constant term in $f_T(z)$. This leads to confinement of the holonomy at two possible values, 0 and $h/2e$.  Sometimes the flux tends to $h/2e$, sometimes to 0, and sometimes both are attractive and the end point depends on the initial conditions.

More surprisingly, on a disk of finite size, we find behavior that drastically differs from the infinite plane. This is perhaps due to the finite size effects we described in section~\ref{nobck}. Specifically, we will find that the flux on a finite disk is quantized to be either $h/2e$ or $0$.
Hence, we will now show that ordinary metals do in fact trap $h/2e$ flux due to the mesoscopic effects!

\subsection{Trapping Flux on a Cylinder}

Let us consider electrons moving on a 2D cylinder. We will explicitly demonstrate that the holonomy exponentially decays (with oscillations) towards two possible values: 0 or $h/2e$. We will use the  coordinates $(\phi,z)$, with $\phi\simeq\phi+2\pi$ and the radius of the cylinder is taken to be $R$. This is the setup of the right panel of Figure~\ref{fig:solenoids}.

\begin{figure}
    \centering
    \includegraphics[width=0.75\linewidth,trim={6cm 10cm 8cm 8cm}, clip]{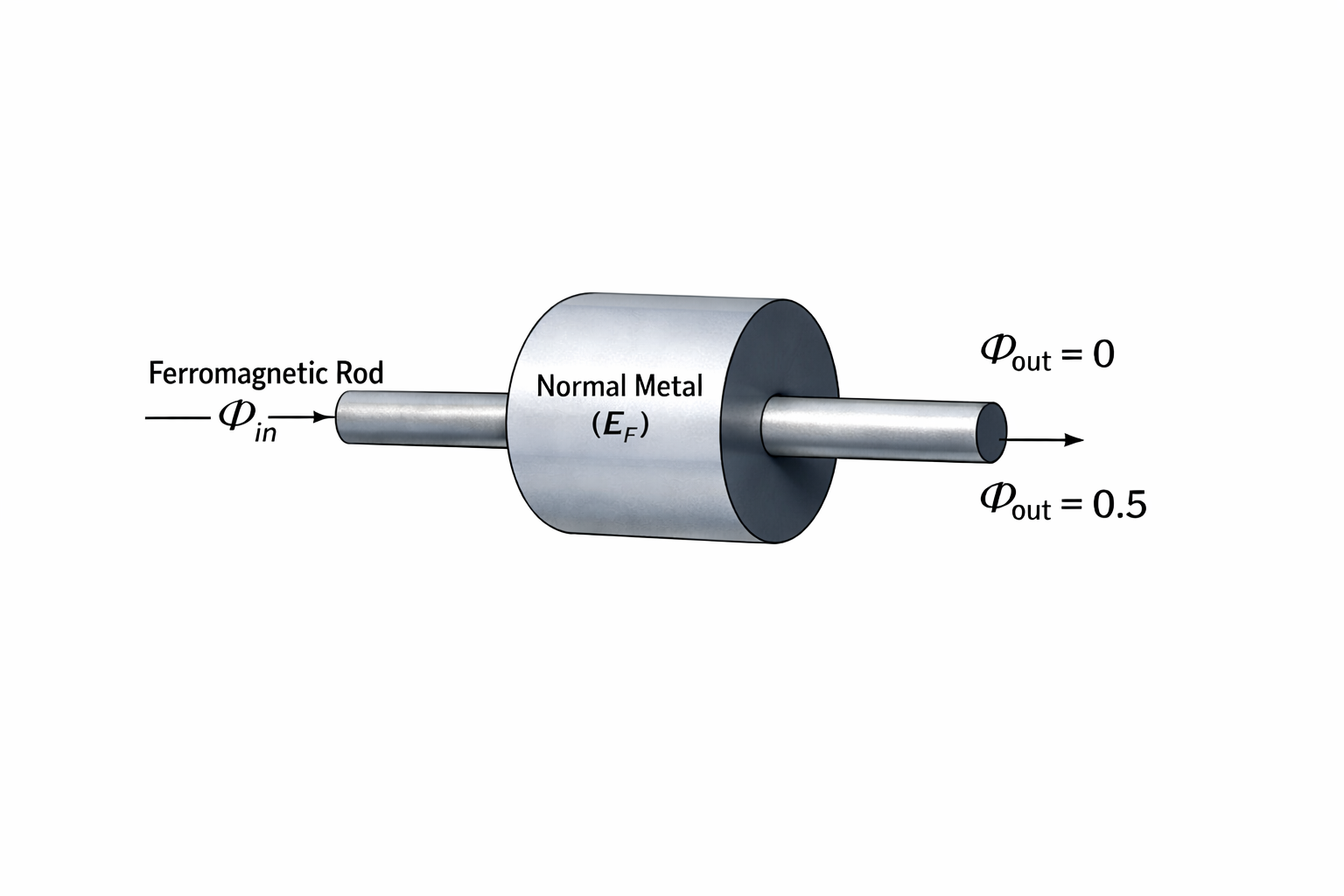}
    \caption{The schematic depiction of a setup, where the flux get trapped with $\Phi_{\rm out}= 0,\tfrac{1}{2}$.}
    \label{fig:cyl_setup}
\end{figure}

Our goal is to minimize the energy
\begin{gather}\label{cylinderproblem}
    E[A_\phi(z), \hat{\Psi}] =  \int   dz d\phi \left[    {1\over \mu_0'R}(\partial_z A_\phi)^2   + \biggl\langle \hat{p}_z^2  + \frac{1}{R^2}(\hat{p}_\phi + A_\phi)^2 - 1\biggr\rangle\right]~,
\end{gather}
the last term in the kinetic energy implements $E_F = 1$. We have set $A_z=0$ without loss of generality. The general two-dimensional Schrödinger equation with arbitrary $A_\phi(z)$ follows from minimizing the second term in~\eqref{cylinderproblem}
\begin{gather}\label{cylgen}
    \left[
        -\,\frac{d^2}{dz^2}
        + \frac{\left(l +A_\phi(z)\right)^2}{R^2}
        \right]
    \psi_{l,k}(z)
    = k^2\,\psi_{l,k}(z),
    \qquad
    k^2 =  E~.
\end{gather}
where the full wave function is $\psi_{l,k}(z,\phi) = e^{-i l \phi} \psi_{l,k}(z)$.
The azimuthal current density generated by occupied one-particle states in the ground state then takes the form
\begin{gather}\label{currgen}
    j_\phi(z)
    = \sum_{l}\sum_{k\le 1}
    \left(l +  A_\phi(z) \right)\,
    \left|\psi_{l,k}(z)\right|^2 ,
\end{gather}
where the sum extends over all states up to the Fermi energy.
First of all, let us note that if we fix  $ A_\phi(z)\equiv \nu$ we would get that the corresponding current is
\begin{gather}
    j_\phi(z) = \frac{1}{\pi R}\sum_{|l +\nu| \leq R} \frac{l
        + \nu}{R} \sqrt{1 - \frac{(l+\nu)^2}{R^2}}~.
\end{gather}
We see that at $\nu=0,1/2$
the current is identically zero for all $R$. For other values of $\nu$ it is generically nonzero, however,
there exist some special values of $R$ where the current is zero for $\nu \neq 0, 1/2$ (for instance, for $
    R = 2.9,  \nu = 0.11858$).
Because for generic  $\nu\neq 0,1/2$ there is a current for all $z$, we should expect a qualitatively significant backreaction from the magnetic energy.

\begin{figure}
    \centering
    \includegraphics[scale = 1.2]{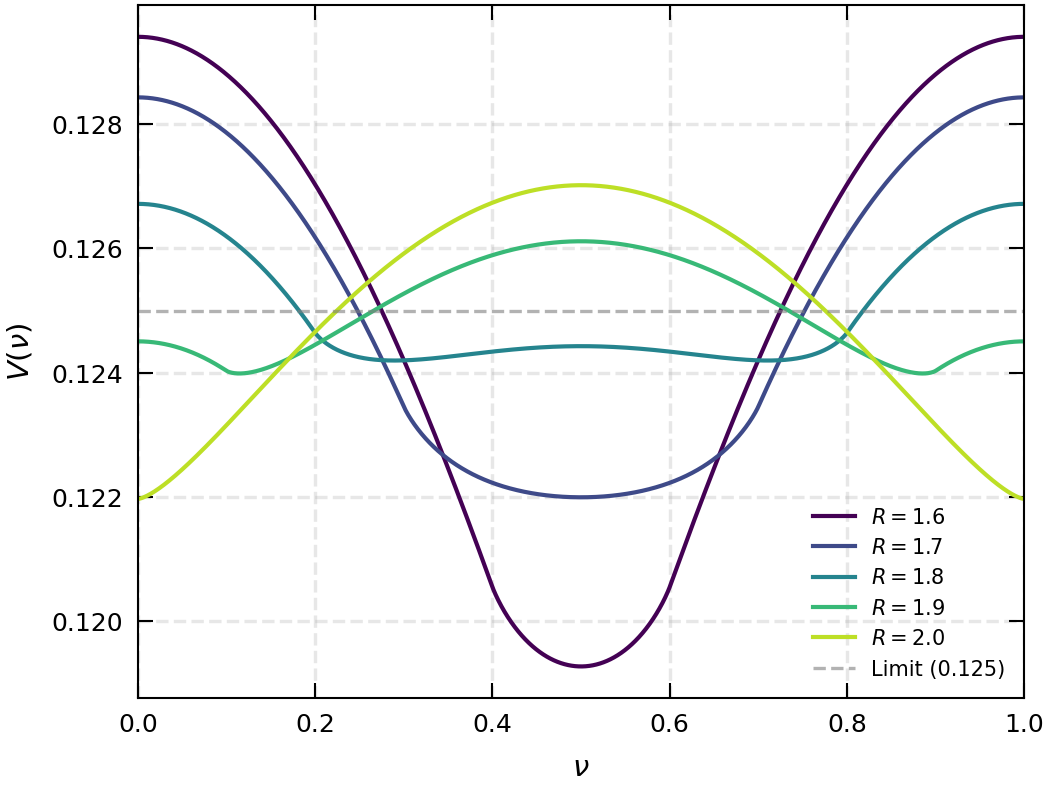}
    \caption{ In the interval $\nu \in [0, 1)$, the potential generally exhibits two extrema located at  $\nu=0$ and $\nu=\frac12$. Whether these points are maxima or minima alternates depending on the radius $R$. At some special values of $R$ the potential also has additional minima at $\nu_*\neq 0,\frac12$.}
    \label{fig:potential}
\end{figure}

To address the full nonlinear problem with the back-reaction we will integrate out the fermions and calculate the exact effective potential for the holonomy $\nu$. We ignore the corrections to the derivative terms since those are inessential for the question of flux trapping.
This approach is well justified for $\mu_0' \ll 1$.
We thus obtain the 2D energy functional
\begin{gather}
    E[A_\phi = \nu] = \int dz  \frac{\pi}{\mu_0'R} \left(\partial_z \nu\right)^2 + \sum_{|l + \nu| \leq R}\int dz\int\limits^{\sqrt{1 - (l+\nu)^2/R^2}}_{-\sqrt{1 - (l+\nu)^2/R^2}} \frac{dk}{2\pi} \left(k^2 - 1\right), \label{magneticenergycylinder}
\end{gather}
Equivalently, the energy density is
\begin{gather}    \mathcal{E}
    = \frac{\pi}{\mu'_0R} (\partial_z \nu)^2 - \Omega(\nu), \quad     \Omega(\nu) = \sum_{|l + \nu| \leq R} \frac{2}{3\pi}\left(1 - \left(\frac{l + \nu}{R}\right)^2\right)^\frac32~.\label{effpot}
\end{gather}
Therefore the holonomy would tend to the local maxima of $\Omega(\nu)$. The same follows from the equation of motion for the holonomy $\nu$
\begin{gather}
    \partial_z^2 \nu = -{\mu'_0R\over \pi}\,\partial_\nu\Omega(\nu), \quad \nu(0) = \nu_0, \quad \nu'(\infty) = 0,
\end{gather}
which shows that $\nu(\infty)$ is a local maximum.
The potentials for various $R$ are depicted in  figure ~\eqref{fig:potential}. We see that either of $\nu = 0,1/2$ is a  local maximum (sporadically, some $\nu_*\neq 0,\frac12$ appears which is a local minimum). Therefore the holonomy is generally attracted to either 0 (trivial) or to $h/2e$. Since the second derivative at those maxima are generically nonzero, the holonomy decays exponentially to these attractors.
The trapping length is estimated from the curvature at these minima, in the limit of large $R$ as
\begin{gather}
    \lambda_{\rm trapping} =  \frac{R^\frac14}{\sqrt{4\pi \alpha k_F^\frac32 \frac{v_F} {c}}}~.
\end{gather}
where we have momentarily restored the physical units to make the length scale easier to read.

As we can see in figure~\ref{fig:potential} whether the holonomy is attracted to the trivial value or to half flux is $R$ dependent. Furthermore, it sometimes happens that both $\nu=0,\frac12$ are attractive and there is a separatrix at some $\nu_*$ which is an unstable equilibrium for the holonomy. Such a $\nu_*$ can exist because for that special value there is ``accidentally" no current.

As a consistency check, and to make contact with the usual RPA approach, it is not too difficult to calculate the corrections to the kinetic term in~\eqref{effpot}. For that we calculate the retarded static current-current correlator.
We set  the angular quantum to zero and consider $\expval{j_\phi(\omega = 0, q) j_\phi(\omega = 0, -q)} = \Pi_{\phi\phi}(q)$. It receives two contributions, one from the loop diagram and one from the contact term
\begin{gather}\label{cylinderret}
    \chi_{\rm para}(q) = \sum_{l}    \frac{(l+\nu)^2}{R^3} \int \frac{dk}{2\pi}  \frac{n((q + k)^2 + (l + \nu)^2/R^2) - n(k^2 + (l + \nu)^2/R^2)}{(q +k)^2 - k^2}~,   \notag\\
    n (\epsilon) = \theta(1 - \epsilon)~, \notag\\
    \chi_{\rm dia}(q) = -\frac{1}{2\pi R} \sum_{l} \sqrt{1 - \frac{( l + \nu)^2}{R^2}}~.
\end{gather}
We can immediately check that at  $q\to 0$ this just reproduces the curvature of the potential~\eqref{effpot}, indeed,
\begin{gather}
    \lim_{q\to 0}\chi_\nu(q) = -\frac{1}{2\pi R }\sum_{|l + \nu| < R } \left[ \sqrt{1- \frac{(l+\nu)^2}{R^2}} -\frac{(l + \nu)^2}{R^2 \sqrt{1- (l+\nu)^2 /R^2}}\right]~.\label{zerom}
\end{gather}
Further expanding in $q$ we can obtain the corrections to the kinetic term.

Note that the constant term in the retarded two-point function must be contrasted with the two-point function on the plane~\eqref{eq:KTstaticmain} which does not have a constant term. This constant term is why the magnetic flux is quantized on the cylinder but not on the plane.

The above trapping of flux can be realized as follows. In the absence of the metal, we can prepare a constant holonomy $\nu$ at all $z>0$ by activating currents along the cylinder localized near $z=0$ as  $J_{\phi,\ \rm ext}(z)\sim \nu \delta'(z)$. But due to the metal, for sufficiently large $z>0$ we now find exponential decay and the holonomy will tend to either $0$ or $h/2e$ depending on the radius $R$. In other words the metal can serve as a device that helps tune the flux to $h/2e$ for generic initial conditions.\footnote{As explained in section~\ref{app:notations} to ignore the tangent components of the magnetic field we can place a material of very high magnetic permeability.
    We also assumed that the metal's transverse thickness is much smaller than $R$.}
Another possible realization is to prepare a  superconducting solenoid with some initial flux $\nu$ at $z=0$. In the absence of a cylindrical metal around it, it is well known that at $\nu=1/2$ and $\nu=0$ the superconducting solenoid has the same energy. The metal we surround it by breaks this degeneracy due to the effective potential~\eqref{effpot}. This should be possible to observe by starting from generic external flux $\nu$ at $z=0$ and letting the system relax towards the ground state at $z\to\infty$.
The geometry of the first setup with a high permeability material everywhere inside the cylinder is depicted in Fig.~\ref{fig:cyl_setup}.

From~\eqref{cylinderret} we can also identify an oscillatory component which means that the exponential decay is accompanied by oscillations of period $\pi$ (in dimensionless units).

\begin{figure}
    \centering
    \includegraphics[width=0.85\linewidth]{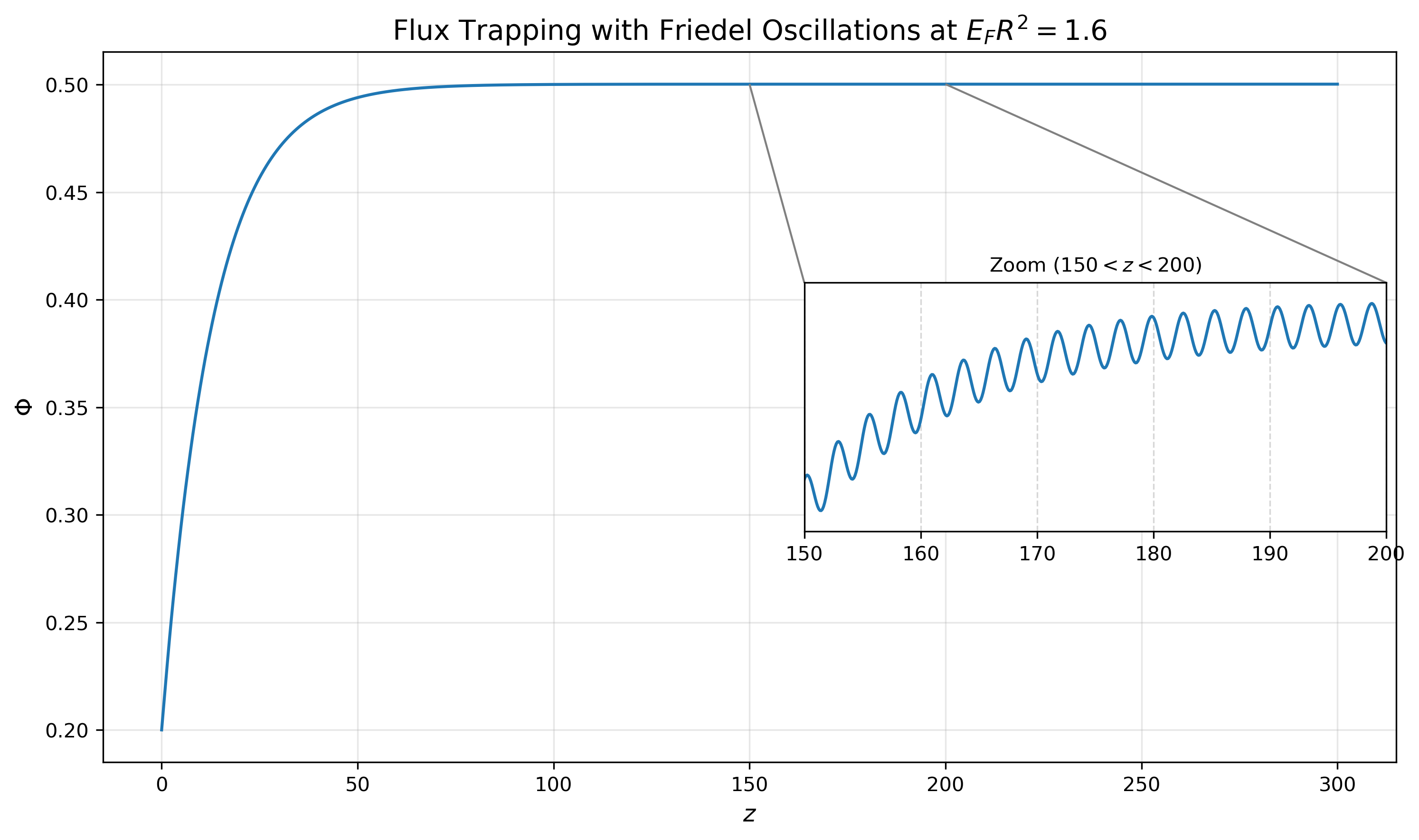}
    \caption{Spatial profile of flux trapping and emergence of Friedel-like oscillations.
        The main panel displays the self-consistent solution for the magnetic flux $\Phi$ as a function of the axial coordinate $z$, calculated for the parameter $E_F R^2 = 1.6$.
        The flux increases monotonically before saturating near the half-flux quantum value.
        The inset highlights the region $150 < z < 200$, revealing Friedel-like oscillations of the flux $\Phi$ as a function of $z$ (with $\lambda_F=\pi$)}
    \label{fig:expattraction}
\end{figure}

We have performed simulations of the full nonlinear problem~\eqref{magneticenergycylinder} to demonstrate the above facts. We numerically solve~\eqref{cylgen} for generic $A_{\phi} (z)$ and then calculate the current~\eqref{currgen} and finally check if \begin{gather}
    \frac{\partial^2 A_\phi}{\partial z^2} = \mu'_0\, j_\phi(z) .
\end{gather}
is obeyed.
If it is not obeyed we perform a step of gradient descent to find a better approximation to the exact solution.
The details are very similar to the procedure we adopt on the disk in the next section so we postpone discussing it to then.
One example of the results we get is in figure~\ref{fig:expattraction} and we zoom in on the oscillations.

\subsection{Trapping $h/2e$ Flux on a Disk}

\begin{figure}
    \centering
    \includegraphics[width=0.5\linewidth]{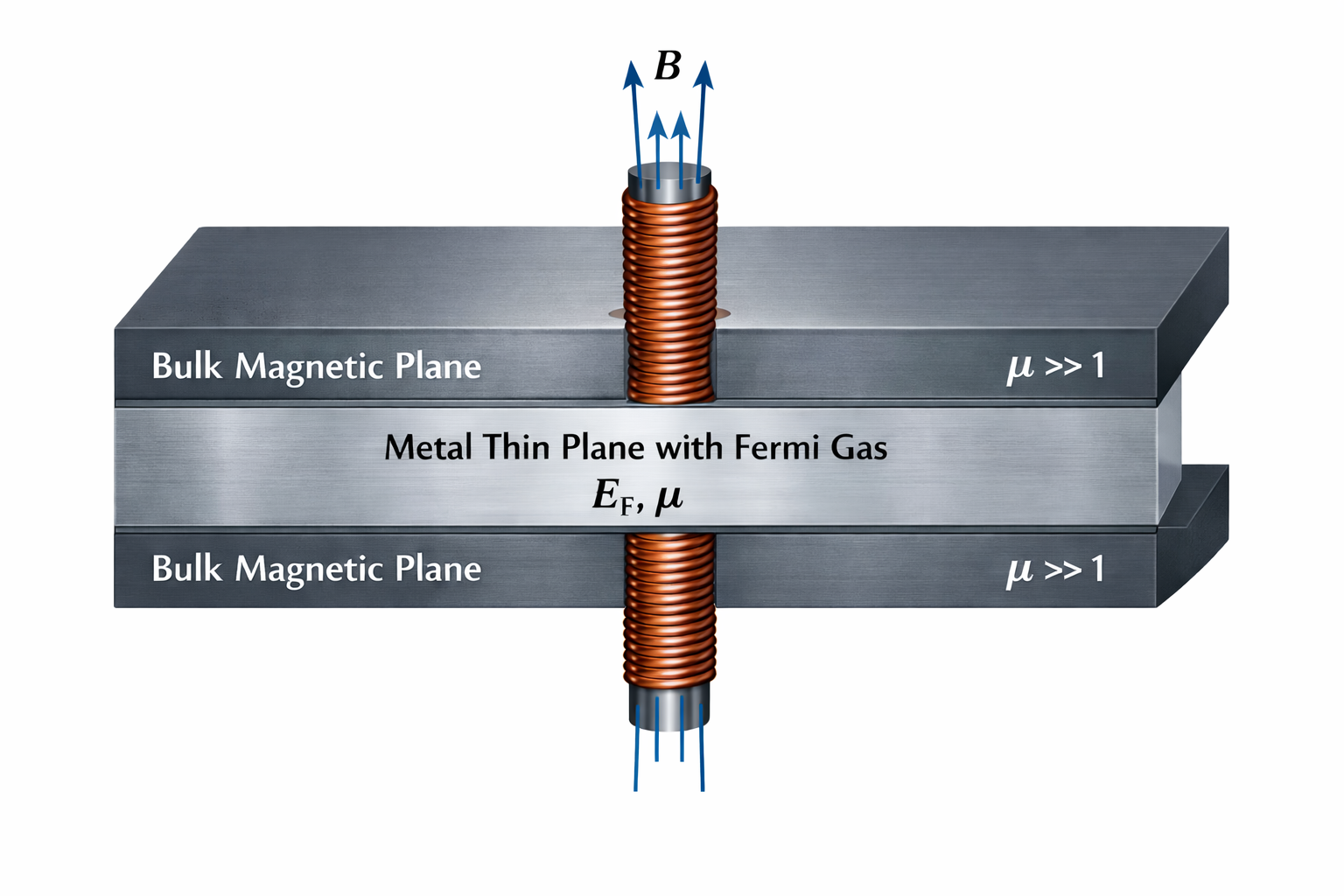}
    \caption{A thin solenoid pierces a normal-metal thin film sandwiched between two bulk high-permeability plates. }
    \label{fig:solwithmu}
\end{figure}

In section~\ref{nobck} we saw that inserting a magnetic solenoid in a plane metal leads to currents which asymptotically decay and oscillate. They create their own magnetic fields via Amp\'ere's law $\vec\nabla\times \vec B_{\rm ind}= \mu'_0 \vec J$.
The magnetic field should then feed back into the electronic wave functions and thus affect the currents. As explained in Section~\ref{RPAreview}, in infinite volume, this backreaction effect is very small and does not change the conclusions in any significant way. There is no intrinsic trapping length scale, in contrast to the situation on the cylinder in the previous subsection.

However, a question remains: what is the total magnetic flux at finite volume? Can it be anything or is it quantized? The infinite volume analysis would suggests that it could be anything  but surprisingly we find  that at finite volume it is quantized to be either 0 or $h/2e$!

Now let us set up the problem precisely.
We solve the coupled Schr\"odinger-Maxwell equations.  In the stationary and axial symmetric setup the following functional is minimized
\begin{gather}\label{diskgeometry}
    E[A_\theta(r), \hat{\Psi}] =   2\pi\int   {dr\over r}    \, \left[\frac{1}{\mu_0'}(\partial_r A_\theta)^2   + \left\langle (\hat{p} - A(\hat{x}))^2 - 1\right\rangle \right]~,
\end{gather}
where the first term corresponds to the energy of the electromagnetic configuration created by the field $A_\theta$. The setup is again such that only the transverse component of the magnetic field contributes, by virtue of the metal being surrounded by a high permeability material.
As before, the $-1$ in~\eqref{diskgeometry} implements $E_F = 1$.
More explicitly, the second term in~\eqref{diskgeometry} is given by
\begin{gather}
    \left\langle (\hat{p} -A(\hat{x}))^2 \right\rangle
    =\sum_{E_{n,p} \leq E_F} \int 2\pi r dr \left[ \left|\partial_r \psi_{n,p}\right|^2 + \frac{\left(n-A_\theta \right)^2}{ r^2}\left| \psi_{n,p} \right|^2\right]~.
\end{gather}
We minimize this functional with respect to $A_\theta$ and $\hat{\Psi}$ with Dirichlet boundary conditions for the wave functions $\psi_{n,m}$ and we fix the holonomy at the inner circle
\begin{equation}\label{holbc}A_\theta(a)=\nu~.\end{equation}We use the same annulus geometry as in section~\ref{nobck} with the inner radius being $a$ and the outer radius being $b$.   To find the minimum energy configuration $A_\theta$ we will follow the following procedure. First we find the electron wave functions with some arbitrary $A_\theta$:
\begin{align}
     & \psi_{n,p} = r^{-\frac12}R_{n,p}(r) e^{in\theta                                                                                                     %
    }~, \qquad \int\limits^b_a dr R_{n,p}(r) R_{n,p'}(r) = \delta_{p,p'}~,  \notag                                                                         \\
     & - \partial_r^2 R_{n,p} + \frac{(n - A_\theta)^2 - \frac14}{r^2} R_{n,p} =  E_{n,p} R_{n,p}~, \qquad R_{n,p}(a) = R_{n,p}(b) = 0~.\label{eq:redSchr}
\end{align}
The ground state corresponds to a state where all energy levels with the energies $E_{n,p} \leq 1$ are filled. We find the  current
\begin{gather}
    j^\theta(r) = \frac{1}{r^3} \sum_{E_{n,p} \leq 1} \left(n - A_\theta\right)\left|R_{n,p}(r)\right|^2~.
\end{gather}
Finally, we check if the Maxwell equation is obeyed
\begin{gather}
    \frac{\delta E}{\delta A_\theta} = \frac{1}{4\pi r} \frac{\partial}{\partial r} \left[ \frac{1}{r} \frac{\partial}{\partial r} A_\theta\right]  -  j^\theta(r)
    \label{eq:grad_step} = 0~.
\end{gather}
If the equation~\eqref{eq:grad_step} is not obeyed then we use gradient descent for $A_\theta$ to find a better approximation of the solution. Namely, we note that if $\lambda$ is a learning rate of our gradient descent then $E\left[A_\theta - \lambda \frac{\delta E}{\delta A_\theta}\right] \approx - \lambda \left( \frac{\delta E}{\delta A_\theta}\right)^2 < 0$. Thus we start with some initial configuration $A_\theta$ and  steadily adjust towards the correct minimum. To solve this problem numerically, we discretize the Schr\"odinger equation along concentric circles of radii $r_i = a + \frac{b-a}{N} i, i =0,\ldots,N$.

\begin{figure}
    \centering
    \hspace*{1cm}\includegraphics[scale=0.2]{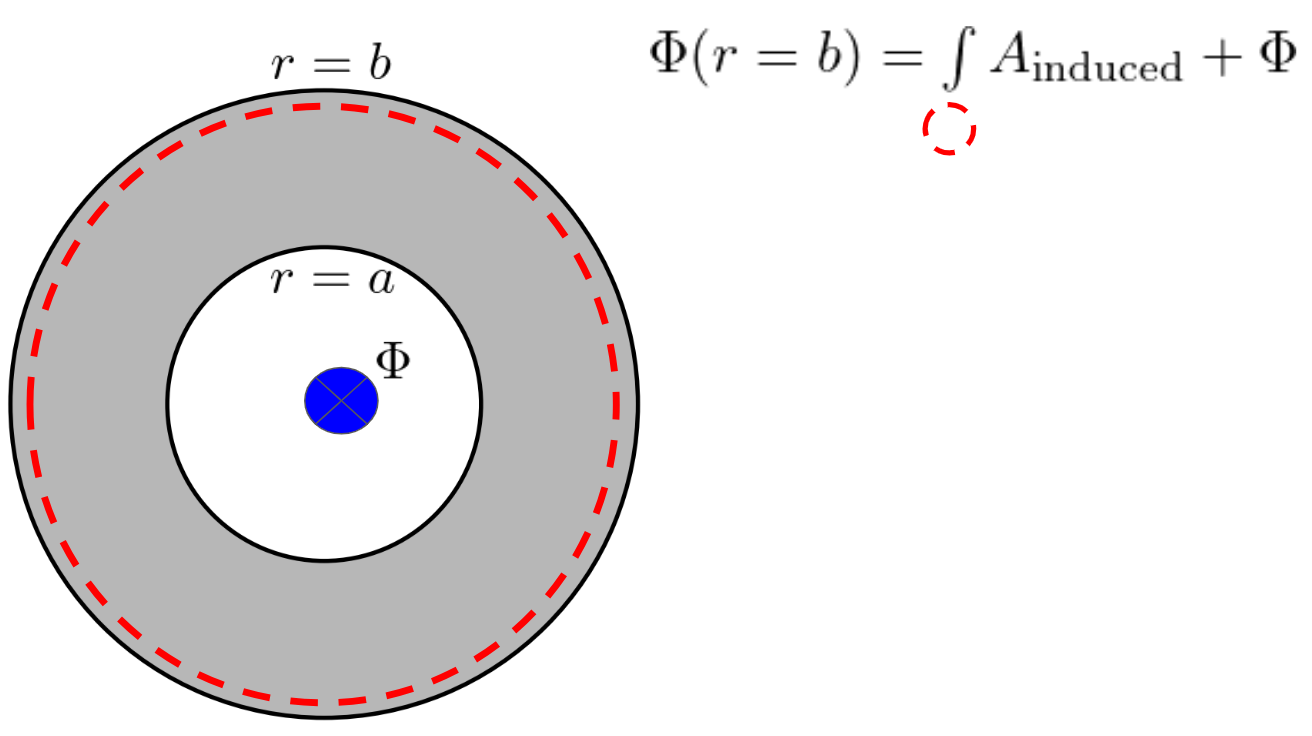}
    \caption{The flux consists of the original flux created by the solenoid and the induced magnetic field due to the metal. }
    \label{fig:quest_scf}
\end{figure}

Finally, once we have a good approximation of the solution we calculate the total magnetic field trapped in the metallic disk
\begin{equation}\Phi(r)=2\pi\int_{r'<r} r'dr' (B_{\rm ind})_z+\Phi~,\end{equation}
(where $\Phi$ is the boundary condition at the inner circle)
where we eventually set $r=b$. See Figure~\ref{fig:quest_scf}.

In relativistic theories in the vacuum it is reasonable to expect that $\Phi/\Phi_0=0$ mod 1 is an attractive fixed point while $\Phi/\Phi_0=\frac12$ mod 1 is a repulsive fixed point.\footnote{In supersymmetric theories, though, the beta function often vanishes exactly, see~\cite{Assel:2015oxa}.} Indeed the repulsive fixed point at $\Phi/\Phi_0=1/2$ played an important role in the recent spin-flux duality~\cite{Komargodski:2025jbu}.
It is surprising that in a Fermi gas the situation is entirely different.
Solving the coupled system of electrons and magnetic fields in the presence of a solenoid we find that  $\Phi(r=b)$ is either $0$ or $h/2e$, leading to flux quantization!
We have speculated in Section~\ref{nobck} that this phenomena occurs due to the presence of large finite-size effects.

\section{Short Distance Behavior and  Perfect Diamagnetism}\label{secanomaly}
In section~\ref{nobck} we already remarked that something interesting is happening for $r<k_F^{-1}$ in the limit that $a\to0$.
The analysis of the boundary conditions as $a\to 0$ is presented in appendix~\ref{bc} along with the renormalization group analysis of the possible boundary conditions.
The most stable fixed point corresponds to modes with angular momentum $\ell\in \mathbb{Z}+\nu$ (with $\nu=\Phi/\Phi_0$) having boundary condition $\Psi \sim r^{|\ell|}$ near the origin. This boundary condition is conformal in the sense that it preserves the Schr\"odinger symmetry algebra .
If we take the $b\to\infty$ limit, i.e. assume that that the metal is very large, we obtain the wave functions
\begin{gather}
    \Psi_{\ell,E} = \sqrt{\frac{k}{2b}} e^{i \ell\theta} J_{|\ell|}(kr)
\end{gather}
and the energy is given by $E=k^2$. %
The azimuthal current then is given by
\begin{gather}
    j^\theta = \sum_{\substack{l \in \mathbb{Z}+\nu\\ E<1}}
    \frac{k\, l}{ b r^2} J^2_{|\ell|}\left(k r\right)
\end{gather}
The sum over $E<1$ has to be interpreted as an integral with the correct density of states. Let us use the identity that $\int\limits^\infty_0 \frac{dr}{r} J^2_\nu(k r) = \frac{1}{2\nu}$ to obtain the total current through the sample
\begin{equation}
    I_{tot} = \int r dr j^\theta = {1\over 2 b}\sum_{\ell} \int_{E=0}^{1} k {dN_\ell\over dE}\operatorname{sgn}(\ell) dE~.\end{equation}
In the large volume limit the density of states ${dN_\ell\over dE}$ is $\ell$ independent and is given by ${dN_\ell\over dE}={b\over 2\pi \sqrt{E}}$ therefore the total current is given by
\begin{equation}
    I_{tot} = {1\over 4\pi}\sum_{\ell} \operatorname{sgn}(\ell)~.\end{equation}
The properly reqularized version of $\sum_\ell\operatorname{sgn}(\ell)$ is given by $1-2\nu$ for $0<\nu<1$ and it is continued periodically past this fundamental domain. Therefore we obtain our final result
\begin{equation} \label{currentzeroa}
    0<\nu<1: \quad   I_{tot} = {1\over 4\pi}\left( 1-2\nu\right)~.
\end{equation}
In particular
\begin{equation}
    I_{tot}(\Phi\to0^+)-I_{tot}(\Phi\to 0^-) ={1\over 2\pi}~.
\end{equation}
The current and the  grand canonical potential are related via $I = -{\partial\Omega\over \partial\Phi}$ which allows us to immediately integrate and obtain at zero temperature
\begin{equation}
    \label{energygs}
    \Omega = {1\over 4\pi}\left(\left|\nu\right|-\nu^2\right), \quad \nu = \frac{\Phi}{\Phi_0}, \quad \left|\nu\right| \leq 1~.
\end{equation}
Due to the linear energy cost of introducing external solenoids with small flux, we interpret this phenomenon as perfect defect-diamagnetism (of course, linear response theory fails due to this non-analyticity).\footnote{A discontinuous derivative of the energy is also characteristic of first-order phase transitions. But then the cusp is a maximum rather than a a minimum. Examples of the former are also known in various defect systems, e.g.   in~\cite{Lanzetta:2025xfw,Zhou:2023fqu,Cuomo:2024psk,Cuomo:2023qvp,Komargodski:2025jbu}. }

\begin{figure}
    \centering
    \includegraphics[width=0.75\linewidth]{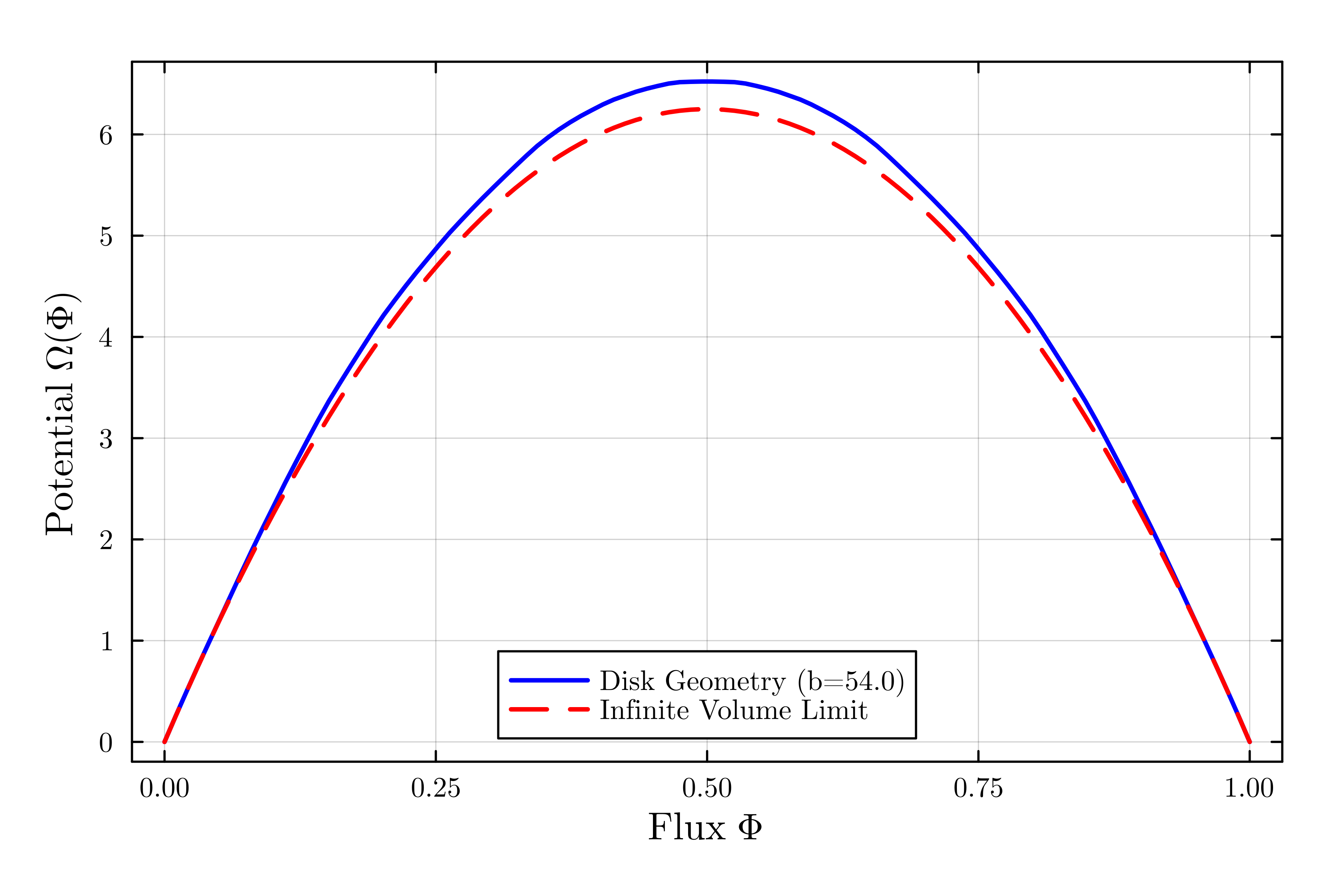}
    \caption{The dependence of grand-canonical potential $\Omega(\Phi)$ on the inserted flux $\Phi$.}
    \label{fig:omega_phi}
\end{figure}

For the sake of completeness, we computed the ground state energy numerically in a large but finite system and in Figure~\ref{fig:omega_phi} we see a nice agreement with the predicted formula \eqref{energygs} .
The grand-canonical potential is non-analytic when $a\to 0$ and it approaches the parabola as we remove the outer boundary.

A key question is what is the spatial profile of the  current~\eqref{currentzeroa}. Of course we have the exact solution for this problem in infinite volume in section~\ref{nobck}. But since the $\nu\to0$ limit is subtle (and in fact non-analytic) this requires extra attention.
For very small $\nu$ the anomalous current is localized near the solenoid. For infinitesimal $\nu$ it is delta function localized. The  solenoid creates  a centrifugal barrier due to which the electronic wave functions near the solenoid behave as $\psi_{m}(r) \sim e^{i(\nu+m) \theta}r^{\nu}$ with $m$ an integer. The $m=0$ modes are the most interesting for our purpose here. Now we use the fact that as a distribution
$\nu r^{-1+2\nu}\sim\delta(r)$ as $\nu\to0$. Therefore, each of the wave functions with $m=0$ gives a nonzero current that remains as the solenoid is removed. Summing over all the $m=0$ modes we get the finite zero-solenoid current above.
In general, the current is mostly concentrated in the region
\begin{equation}\label{exponent}
    r \lesssim  e^{-{1\over \nu}}~.
\end{equation}
Let us briefly comment on the stability of this phenomenon:
\begin{enumerate}
    \item  As we emphasized in the introduction this phenomenon requires control over physics on short distance scales~\eqref{exponent}. It is really just a property of the Fermi gas. Realistically, we can consider situations where the scale $k_F^{-1}$ is much larger than the lattice scale or the scale where electron cease to behave as effectively-free particles. Semi-metals could be reasonable candidates. Under such circumstances, the large diamagnetism could be observable.

    \item Large diamagnetism is stable to disorder since it is mostly a consequence of the centrifugal force near the solenoid. \item In~\eqref{energygs}
          we have ignored the back-reaction of the currents. This cannot remove the discontinuity of the derivative of the grand canonical potential, so this approximation is justified.
\end{enumerate}

From~\eqref{exponent} we can infer what happens when $a$ is small but nonzero.
The distance scale $a$ can represent the solenoid's width, or the inner circle of the metal annulus.
We expect that a large current remains as long as $|\log( a)|\geq \nu $.

At finite small $a$
we observe a crossover phenomenon due to anomalously large currents at the inner edge while at $a\to0$ we observe non analyticity.
We will now exhibit the crossover phenomenon at finite small $a$ quantitatively.

In the case of the annulus geometry we have the following equation for the eigenmodes
\begin{gather}\label{annulusev}
    J_\ell(k_{\ell;n} a) J_{-\ell}(k_{\ell;n} b) - J_\ell(k_{\ell;n}b) J_{-\ell}(k_{\ell;n
        }a) = 0~.
\end{gather}
We are interested in the behavior at small $\nu$.
We concentrate on the $\ell=\nu\ll1$ modes. We   assume a narrow solenoid (or narrow annulus) $a\ll1$ and we also take the outer radius to be large $ b\gg1$.
We get the following equation for the eigenvalues
\begin{gather}\label{eigenexpansion}
    \frac{2^{-\nu } (k_{\nu; n} a)^{\nu } \cos \left( k_{\nu; n} b - \frac{1}{4} \pi  (1 - 2 \nu)\right)}{\Gamma (1 + \nu)}-\frac{2^{\nu } (a k_{\nu; n})^{-\nu } \cos \left( k_{\nu; n} b - \frac{1}{4}
        \pi  (1 + 2 \nu )\right)}{\Gamma (1-\nu )} = 0
\end{gather}
That allows us to find that
\begin{gather}
    k_{\nu;n} b \approx -\frac{\pi}{4} + \pi n+ \frac{\pi}{2\log\left(\frac{\pi n a}{2 b}\right)} - \frac{\pi \nu^2}{6} \log\left(\frac{\pi n a}{b}\right)+  \mathcal{O}(\nu^4)~,
\end{gather}
where we truncated the expansion at second order in $\nu$ and the leading terms at large $n$.
We now need to fill these modes to the Fermi energy. The grand canonical potential is therefore
\begin{equation}\label{gcfinite}
    \Omega={\rm const} - \frac{\pi^2 \nu^2}{12 b^2} \sum_{n\lesssim n_{\rm max} = { b\over \pi}}n \log\left(\frac{\pi n a}{ b}\right) +  \mathcal{O}(\nu^4)~,
\end{equation}
Finally, estimating the above sum we obtain
\begin{equation}\label{gcfinite}
    \Omega={\rm const}-\frac{1}{24}\nu^2 \log\left(a\right)+\ldots~.
\end{equation}
We see that for finite $a$ the non-analyticity turns into a crossover with a large second derivative at the origin
$\frac{\partial^2 \Omega}{\partial \nu^2} = -\frac{1}{12} \log\left( a\right) > 0$.
We can think of~\eqref{gcfinite} as  log-enhanced diamagnetism.

\begin{figure}
    \centering

    \begin{subfigure}{0.32\textwidth}
        \includegraphics[width=\linewidth]{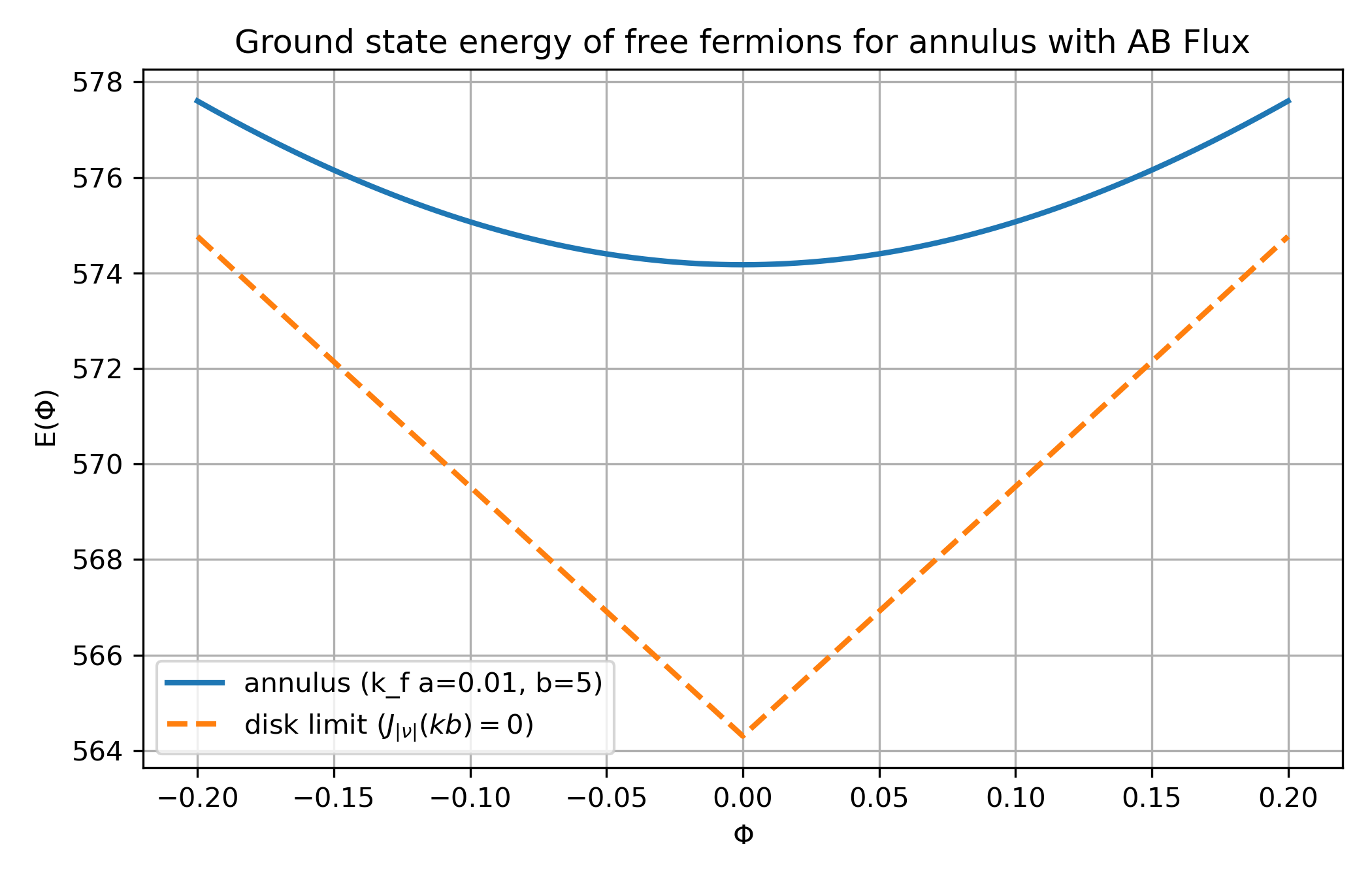}
        \caption{$k_F a=10^{-2}, b=5$}
    \end{subfigure}
    \begin{subfigure}{0.32\textwidth}
        \includegraphics[width=\linewidth]{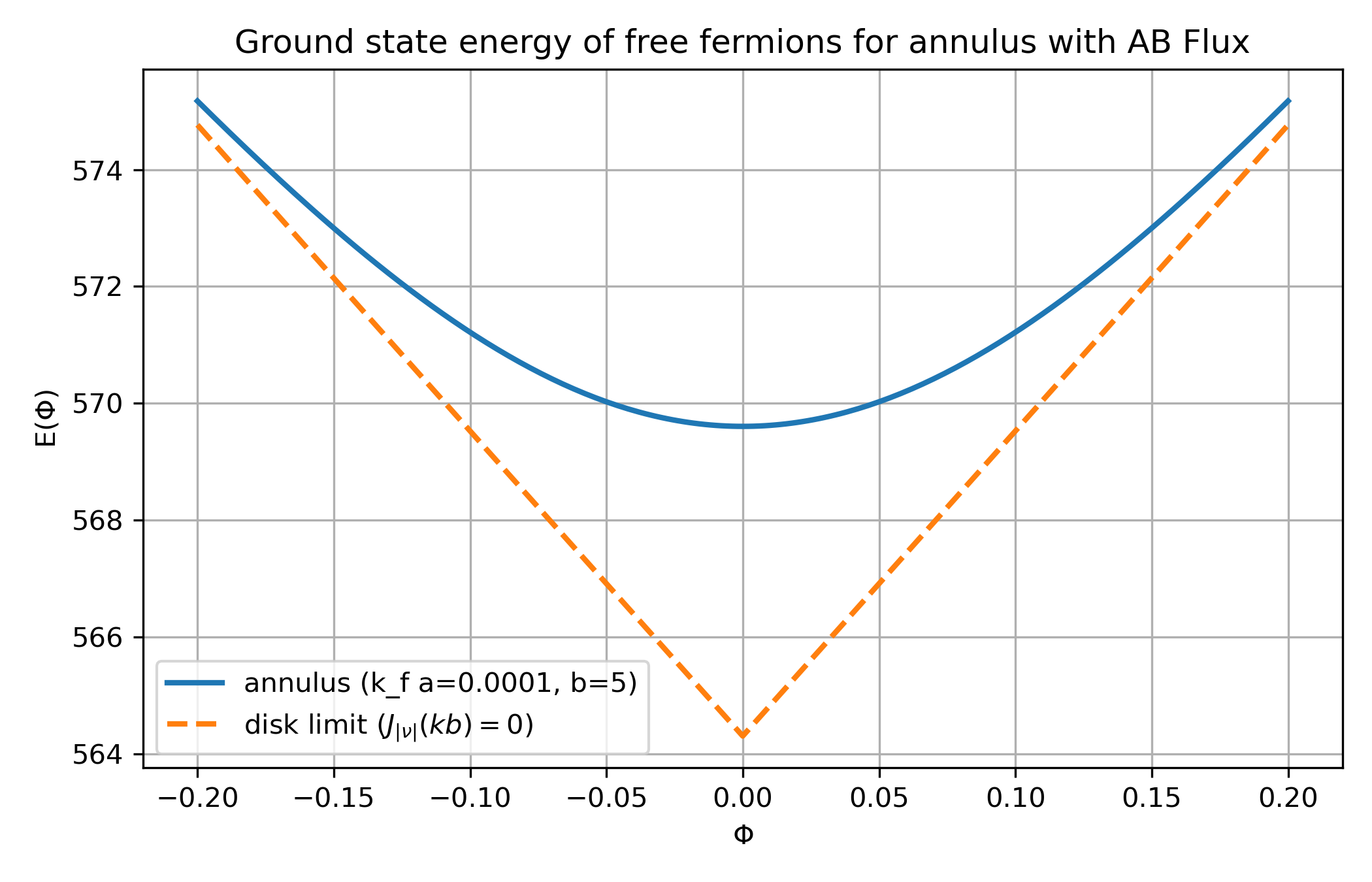}
        \caption{$k_F a=10^{-4}, b=5$}
    \end{subfigure}
    \begin{subfigure}{0.32\textwidth}
        \includegraphics[width=\linewidth]{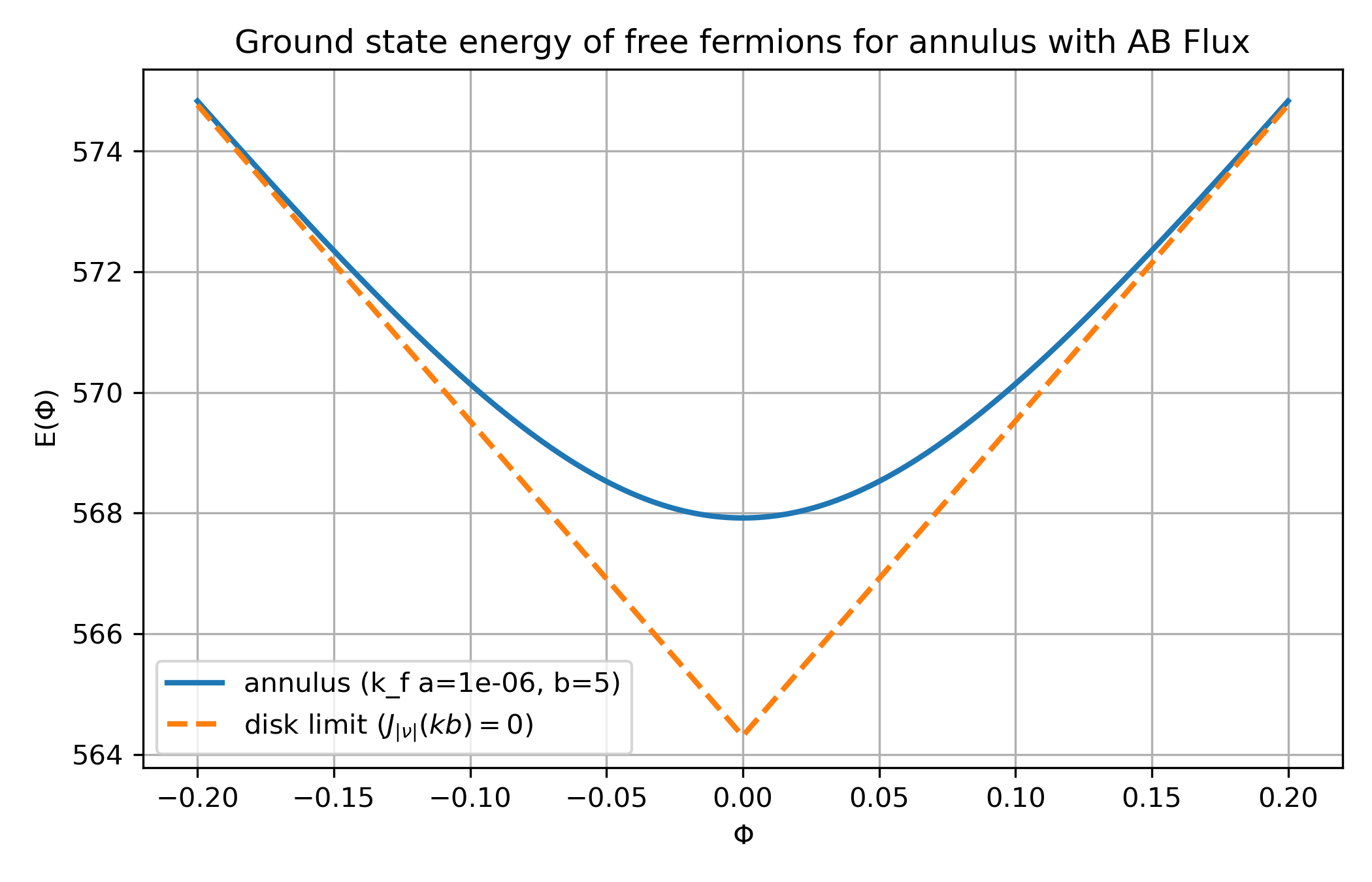}
        \caption{$k_Fa=10^{-6}, b=5$}
    \end{subfigure}

    \caption{The ground state energy of a an annulus with inner radius $r = a$ and outer radius $r = b$ as a function of the solenoid flux $\nu={\Phi\over\Phi_0}$. The solid line is the answer obtained with the parameters specified below the figure and the dashed line is the result at the point $a=0$ where we have non-analyticity.}
    \label{fig:sum_eigenvalues}
\end{figure}

To conclude, we demonstrate the results numerically  in Figure~\ref{fig:sum_eigenvalues}. We see that in accordance with the estimate~\eqref{exponent} an anomalously large current appears for $\Phi \sim |\log(k_F a)|\Phi_0 $. If one can prepare systems with $k_F a\sim 10^{-3}-10^{-4}$ a very large current would appear already for $\Phi/\Phi_0\sim 0.05-0.1$.

\section{Acknowledgment}

We thank Alexander Abanov, Grisha Falkovich, Patrick Lee, Leonid Levitov, Andy Lucas, Subir Sachdev, Daniil Antonenko and Yifan Wang for valuable
discussions. We thank Brandon Rayhaun for his involvement in the early stages of this work. Z.K. gratefully acknowledges
NSF Award Number 2310283. F.K.P. gratefully acknowledges support from the Theoretical Sciences Visiting Program (TSVP) at the Okinawa Institute of Science and Technology (OIST), during which the final part of this research was conducted.

\appendix

\section{Boundary Conditions at the Solenoid}\label{bc}

In the spirit of the renormalization group it is useful to start with a solenoid that has a vanishingly small cross-section. We need to understand the boundary conditions on the electronic wave functions. These boundary conditions should be thought as a self-adjoint extension of our Hamiltonian in the presence of the solenoid.

We will consider electrons in 2+1 dimensions moving in a local axisymmetric potential \begin{equation}\label{potential}V(r)=U\delta^{(2)}(r)+U'\nabla^2\delta^{(2)}(r)+\cdots\end{equation} around a solenoid.
If we write the wave functions as $\Psi(r,\theta) = {1\over \sqrt r} e^{i\ell \theta}\Psi_\ell(r)$ then the eigenfunctions are subject to the equation
\begin{equation}\label{QMcontact}
    \left[-{1\over 2m}\partial_r^2+{\ell^2-\frac14\over 2m r^2}+V(r) \right]\Psi_{\ell} = E\Psi_{\ell}~,
\end{equation}
with $\ell\in \mathbb{Z}+\Phi/\Phi_0$.

Near the origin the wave function behave as $\Psi_{\ell} \sim r^{\lambda}$ with
$\lambda(\lambda-1) = \ell^2-\frac14$. Requiring that the wave function is normalizable implies that for $|\ell|\geq 1$ there is a unique solution which is admissible. For generic values of $\Phi$ there are two modes with $|\ell|< 1$ and their boundary conditions should be determined to make the Hamiltonian self-adjoint.

Assuming without loss of generality that $\nu\equiv\Phi/\Phi_0\in(0,1)$ let us first discus the mode $\ell=\nu$.
We regularize the solenoid core by an axisymmetric potential occupying $0<r<a$ with height $\tilde U$. It is now straightforward to calculate the phase shift.
The solution for $r>a$ is a linear combination $\Psi \sim \sqrt{r}\left( J_\nu(r\sqrt {2mE})+B Y_\nu(r\sqrt {2mE})\right) $. At low energies we find
\begin{align} &B^{-1}={2^{2\nu}\Gamma(\nu)\Gamma(1+\nu)\over \pi}{ \sqrt{-\pi U}J'_\nu(\sqrt{-\pi U})  + \nu J_\nu(\sqrt{-\pi  U})\over  \sqrt{-\pi U}J'_\nu(\sqrt{-\pi U})  - \nu J_\nu(\sqrt{-\pi  U}) }(R\sqrt E)^{-2\nu}-\cot(\pi \nu) ~.\end{align}
Importantly, what appears in the formula above is $U=\pi a^2\tilde U$ which we keep finite in the point-like limit of the solenoid. The result simplifies if we define
$$\nu+{\pi u\over 2(1+\nu)} \equiv {\sqrt{-\pi U}J'_\nu(\sqrt{-\pi U}) \over J_\nu(\sqrt{-\pi  U})}~, $$
such that for very small $u= U+\cdots$. In terms of this variable
$$B^{-1}={2^{2\nu+1}\Gamma(\nu)\Gamma(1+\nu)\over \pi}{ 2\nu(1+\nu)+{\pi u\over 2(1+\nu)}  \over  \pi u }(a\sqrt E)^{-2\nu}-\cot(\pi \nu) $$
If we require that the value of the coupling $u$ evolves in such a way that the $R$ dependence cancels out (so that the thin solenoid limit makes sense) then we obtain a simple renormalization group  equation
\begin{equation}\label{RGeq}{d\over d\log a}u =-2\nu u-{\pi \over 2(1+\nu)^2}u^2 ~.\end{equation}
There are two fixed points of the renormalization group with $u=0$ being the infrared stable fixed point and a fine-tuned fixed point with $u<0$. These are conformal quantum systems and they can be upgraded to full-fledged Schr\"odinger fixed points in $2+1$ dimensions.\footnote{It is worth noting~\eqref{RGeq} that without an Aharonov-Bohm solenoid the only fixed point is $u=0$ and there is no fixed point with an attractive potential. Furthermore, the $u=0$ fixed point is trivial at $\nu=0$, e.g. it has a trivial phase shift. }

If we focus on the attractive fixed point with $u=0$ then $B\to 0$ and the wave functions are purely $J$ modes. That means that the defect interaction  $|\Psi|^2$ (which corresponds to the coupling $U$ in~\eqref{potential}) on the solenoid is irrelevant with
$\Delta(|\Psi|^2)=2+2\nu$.\footnote{In the infrared unstable fixed point $B^{-1}=-\cot(\pi \nu)$ at low energies and corrections to the phase shift behave like $\sim (R\sqrt E)^{-2\nu}$ which grow in the low energy or large cutoff limit, signaling that there is a relevant operator.
    The wave function near the origin consists of the $r^{-\nu}$ mode without the $r^{\nu}$ mode. We can therefore infer that the scaling dimension of the contact interaction $|\Psi|^2$ is  $\Delta(|\Psi|^2)=2-2\nu$.
}

Now let us discuss the mode $\ell=-1+\nu$, which also has two possible boundary conditions. The analysis is identical, though we can identify the corresponding interaction with the defect operator $\int dt \nabla^2\Psi^\dagger \Psi$ -- and we find that it has the scaling dimensions at the two fixed points $$\Delta =4-2\nu~,\quad \Delta = 2\nu~. $$
(The defect operator $\int dt \nabla^2\Psi^\dagger \Psi$ corresponds to the continuum parameter $U'$ in~\eqref{potential}.)

As we increase $\nu$ towards $\nu\to 1^{-}$, we see that the two couplings switch -- and that is crucial for the periodicity of the problem.

There are two conclusions to this discussion:
\begin{itemize}
    \item The non-fine tuned boundary conditions for electrons interacting with the Aharonov-Bohm solenoid require to pick, at $r\to 0$,  $r^\nu$ and $r^{1-\nu}$ for the modes $\ell=\Phi/\Phi_0$ and $\ell=-1+\Phi/\Phi_0$.
    \item The usual contact interaction in 2+1 dimensions has only one fixed point, which is trivial, but here we see that with an Aharanov-Bohm solenoid there is a rich set of fixed points. These fixed points are analogous to the fermions at unitarity fixed point in 3+1 dimensions.

\end{itemize}

\section{Electric and Magnetic RPA calculations}\label{RPAappendix}

We consider a uniform, non-interacting, spinless electron gas in three spatial dimensions at zero temperature,
with dispersion
\begin{equation}
    \varepsilon_{\bm k}=\frac{\hbar^2 k^2}{2m},\qquad k=\abs{\bm k},
\end{equation}
and ground state Fermi sea with Fermi wavevector $k_F$ (so $\varepsilon_F=\hbar^2 k_F^2/(2m)$).
The unperturbed occupation numbers are
\begin{equation}
    n_{\bm k}=\Theta(k_F-k)\qquad (T=0).
\end{equation}
The density of states in the window $dk$ is $dn={1\over 2\pi^2}k^2dk$ and hence the total density
$n={1\over 6\pi^2}\left({2mE_F\over \hbar^2 }\right)^{3\over 2}$. We can also write the total density as $n=\int {d^3k\over (2\pi)^3}\Theta(k_F-k)$.

Let us briefly review the response theory to an external potential $\phi(\bm r)$: \begin{equation}
    H = \int d^3r\;
    \psi^\dagger(\bm r)
    \frac{-\hbar^2}{2m}\nabla^2
    \psi(\bm r)
    -\;e\int d^3r\,\phi(\bm r)\,\hat n(\bm r),
    \qquad \hat n(\bm r)=\psi^\dagger(\bm r)\psi(\bm r).
\end{equation}
Our goal is to compute the induced density perturbation
$\delta n(\bm r)=\expval{\hat n(\bm r)}_\phi-\expval{\hat n(\bm r)}_{\phi=0}$.
(There is also a constant shift of $H$ due to the positive non-dynamical background charge that makes the system neutral as a whole.)

As is standard in linear response theory, we need the retarded Green function
\begin{equation}
    \chi(\bm q,t)= -\frac{i}{\hbar}\Theta(t)\,\expval{\comm{\hat n_{\bm q}(t)}{\hat n_{-\bm q}(0)}}_0.
\end{equation}
Inserting a complete set of particle-hole states we obtain
\begin{align}
    \chi(\bm q,t)
     & = -\frac{i}{\hbar}\Theta(t)\sum_m
    \left[
        \abs{\bra{m}\hat n_{\bm q}\ket{0}}^2 e^{-i(E_m-E_0)t/\hbar}
        -\abs{\bra{m}\hat n_{-\bm q}\ket{0}}^2 e^{+i(E_m-E_0)t/\hbar}
        \right].
\end{align}
Fourier transform in time and use $\int_0^\infty dt\,e^{i(\omega-\omega_{m0}+i0^+)t}= \frac{i}{\omega-\omega_{m0}+i0^+}$,
with $\omega_{m0}=(E_m-E_0)/\hbar$, one obtains
\begin{equation}
    \chi(\bm q,\omega)=\frac{1}{\hbar}\sum_m
    \left[
        \frac{\abs{\bra{m}\hat n_{\bm q}\ket{0}}^2}{\omega-\omega_{m0}+i0^+}
        -\frac{\abs{\bra{m}\hat n_{-\bm q}\ket{0}}^2}{\omega+\omega_{m0}+i0^+}
        \right].
    \label{eq:lehmann}
\end{equation}

The ground state is the filled Fermi sea
$\ket{0}=\prod_{k<k_F} c^\dagger_{\bm k}\ket{\mathrm{vac}}$.
A one particle--hole excitation is $c^\dagger_{\bm k+\bm q}\,c_{\bm k}\ket{0}$
with excitation energy $\varepsilon_{\bm k+\bm q}-\varepsilon_{\bm k}$.
One checks that for an allowed particle--hole pair,
$\abs{\bra{m}\hat n_{\bm q}\ket{0}}^2=1$ (and is zero otherwise). Therefore \eqref{eq:lehmann} becomes
\begin{equation}
    \chi_0(\bm q,\omega)
    =
    \sum_{\bm k}\frac{n_{\bm k}-n_{\bm k+\bm q}}{\hbar\omega+\varepsilon_{\bm k}-\varepsilon_{\bm k+\bm q}+i0^+}\to
    \int\frac{d^3k}{(2\pi)^3}\,
    \frac{n_{\bm k}-n_{\bm k+\bm q}}{\hbar\omega+\varepsilon_{\bm k}-\varepsilon_{\bm k+\bm q}+i0^+}.
    \label{eq:lindhard_sum_box}
\end{equation}
where we switched to the thermodynamic limit.

We evaluate the integral in the static limit $\omega=0$ to obtain an explicit result
\begin{align}
    \chi_0(\bm q,0)
     & =-\frac{2m}{\hbar^2}\int_{k<k_F}\frac{d^3k}{(2\pi)^3}\,
    \frac{1}{q^2+2\bm k\cdot \bm q}
    +\frac{1}{q^2-2\bm k\cdot \bm q}
    \\ & =-\frac{m}{4\pi^2\hbar^2}\biggl[
        k_F+\frac{4k_F^2-q^2}{4q}\ln\abs{\frac{2k_F+q}{2k_F-q}}\biggr]= -\frac{m k_F}{2\pi^2\hbar^2}
    + \frac{m}{24\pi^2\hbar^2 k_F}\,q^2
    +\cdots~.
\end{align}
This is the famous Lindhard function in 3D.
Famously, the first term can be interpreted as $-\nu(E_F)$, i.e. the density of states near the Fermi energy. Indeed, ${dn\over dE}\bigr|_{E=E_F}={mk_F\over 2\pi^2\hbar^2}  $.
In linear response
$\delta n(\bm q)=e\chi_0(\bm q)\,\phi(\bm q)$.

The density perturbations generate their own electric fields and it is interesting to take that into account. A self consistent ansatz, the ``Random Phase Approximation (RPA),'' is where we keep bubble chains and drop everything else.
More precisely, we add a classical term to the energy, $\int d^3r \frac1{8\pi}(\grad \phi)^2$ so our full energy function reads
\begin{equation}\label{RPA}
    \mathcal{E}= \expval{ \int d^3r\;
        \psi^\dagger(\bm r)
        \frac{-\hbar^2}{2m}\nabla^2
        \psi(\bm r)
        \;-\;e\int d^3r\,\phi(\bm r)\,\hat n(\bm r)}+\int d^3 r \ n_{\rm ext}(\bm r)\phi(\bm r)+\int d^3r{1\over 8\pi} (\grad \phi)^2~.
\end{equation}
Here, $n_{\rm ext}(\bm r)$ is a prescribed charge density that is non-dynamical (external). It includes the constant charge density that makes the whole system neutral, plus possible additional external charges.
Most importantly, the electric potential is treated non-dynamically.

Within the leading RPA approximation we calculate the expectation value to second order in $\phi$ using our formula in linear response theory, to obtain
\begin{equation}\label{RPA}
    \mathcal{E}= \int {d^3 q\over (2\pi)^3}\left[  n_{\rm ext}(\bm q)\phi(-\bm q)+{q^2-4\pi e^2 \chi_0(q)\over 8\pi}|\phi(\bm q)|^2~.\right]
\end{equation}
At very small momentum we can just use $\chi_0(q)=-\nu(E_F)$ to see that the electric potential is essentially massive in a metal. This is of course the familiar statement that electric fields cannot penetrate a metal.

For a point impurity of charge $n_{\rm ext}=Qe\delta^{(3)}(r)$ we obtain
$$\phi(q) = {4\pi Qe/q^2\over-1+4\pi e^2 \chi_0/q^2}~. $$
This leads to an exponential decay at long distances with $e^{-r k_{TF}}$, where $k^2_{TF} = 4\pi  e^2\nu(E_F)$.
The singularity at momentum $q=2k_F$ leads to a long-distance oscillatory tail,
\begin{align}
    \delta n(r)
     & \sim
    \frac{\cos(2k_F r)}{r^3}.
\end{align}
(everywhere above $q_0=2k_F$).
These are the Friedel oscillation in 3D.

We now to repeat the analysis for magnetic sources, i.e. currents.
We work in Coulomb gauge $\nabla\cdot\mathbf A=0$.
The energy functional is \begin{align}
    \mathcal E[\mathbf A]
     & =
    \expval{
        \int d^3r\;
        \psi^\dagger(\mathbf r)\,
        \frac{1}{2m}\Big(-i\hbar\nabla+\frac{e}{c}\mathbf A(\mathbf r)\Big)^2
        \psi(\mathbf r)}
    \\ &-\frac{1}{c}\int d^3r\,\mathbf J_{\rm ext}(\mathbf r)\cdot\mathbf A(\mathbf r)
    +\int d^3r\,\frac{1}{8\pi}\,(\nabla\times\mathbf A)^2
    \label{eq:minimal}
\end{align}
We can write the current in the standard decomposition
\begin{align}
    j(\bm r)
    =
    j_p(\bm r)
    +j_{d}(\bm r),
\end{align}
where
\begin{align}
    j_p(\bm r)= &
    -\frac{e\hbar}{2mi}
    \Big(
    \psi^\dagger(\mathbf r)\,\nabla\psi(\mathbf r)
    -
    (\nabla\psi^\dagger(\mathbf r))\,\psi(\mathbf r)
    \Big), \quad
    j_{d}(\bm r)
    =
    -\frac{e^2}{mc}\,\hat n(\mathbf r)\,\mathbf A(\mathbf r).
\end{align}
The first term is often called paramagnetic and the latter diamagnetic.

To the leading approximation, we only need the Retarded two point function of $j_p$ and the one point function of $j_d$, which is the density.
We therefore need to calcualte \begin{align}
    \chi_{ij}^R(\bm q,\omega)
    = -i\int_0^\infty dt\;e^{i(\omega+i0^+)t}\,
    \ev{[\,j_{p,i}(\bm q,t),j_{p,j}(-\bm q,0)\,]}_{0}.
\end{align}
The final answer includes the seagull contact term from $j_d$, so we get
\begin{align}
    K_{ij}^R(\bm q,\omega)=\chi_{ij}^R(\bm q,\omega)+\frac{ne^2}{m}\,\delta_{ij},
    \qquad
    \ev{j_i(\bm q,\omega)} = -\frac{1}{c}K_{ij}^R(\bm q,\omega)\,A_j(\bm q,\omega),
\end{align}
where $n(\bm r)=\ev{n(\bm r)}$ is the average density.

While $\chi_{ij}^R$ in general has transverse and logitudinal components, \begin{align}
    \chi_{ij}^R(\bm q,\omega)
    = P^T_{ij}(\hat{\bm q})\,\chi_T^R(q,\omega)+P^L_{ij}(\hat{\bm q})\,\chi_L^R(q,\omega),
    \qquad
    P^T_{ij}=\delta_{ij}-\frac{q_iq_j}{q^2},\quad
    P^L_{ij}=\frac{q_iq_j}{q^2}.
\end{align}
The physical two-point function $K_{ij}^R$ is fully transverse by current conservation (gauge invariance). This is exactly what the Seagull term achieves. Therefore
$\chi_L^R(q,0)= -\frac{ne^2}{m}$.

Below we use $n=\frac{k_F^3}{6\pi^2}$ and also $z\equiv \frac{q}{2k_F}$.
One finds after a short calculation that we review below
\begin{align}\
    K_T^R(q,0)=\frac{ne^2}{m}\,f_T(z),
    \qquad
    f_T(z)=\frac{3}{8}\qty(1+z^2)-\frac{3}{16z}\,(1-z^2)^2\ln\abs{\frac{1+z}{1-z}}.
    \label{eq:KTstatic}
\end{align}
This is the \emph{full} (gauge-invariant) retarded Green function.
(Equivalently, $\chi_T^R(q,0)
    =\frac{ne^2}{m}\,\qty[f_T(z)-1]$.)

In transverse gauge we therefore obtain the energy
\begin{align}
    \mathcal E[\mathbf A]
     & =
    \int {d^3q\over (2\pi)^3}\left[ -\frac{1}{c}\mathbf J_{\rm ext}(\mathbf r)\cdot\mathbf A(\mathbf r)
    +{1\over 8\pi} q^2|\bm A(\bm q)|^2\left(1+{ne^2\over mc^2}{f_T(z)\over q^2}\right)\right]~.
\end{align}
Since for $z\ll 1$, $K_T^R(q,0)=\frac{ne^2}{m}\left(\frac{q^2}{4k_F^2}-\frac{q^4}{80k_F^4}+ \cdots\right)$,
there is no screening of magnetic fields and the RPA at infinite volume shows that the classical source is slightly corrected.
Note that using $n=k_F^3/(6\pi^2)$ one can read out the standard Landau diamagnetic susceptibility
\begin{align}
    \chi_{\mathrm{Landau}}=-\frac{e^2 k_F}{24\pi^2 m c^2}~.
\end{align}
from the fact that the energy to establish the field goes up $\Delta \mathcal E[\mathbf A]_{matter} = - \chi_{\mathrm{Landau}} B^2$. But indeed, as is well known, there is no screening of the external magnetic field.

The above formalism gives us an important insight into the non analytic dependence of the ground state energy.
Consider a solenoid oriented in the $z$ direction with magnetic flux $\theta$.  That corresponds to the gauge field configuration
$A_i(\bm q)\sim \epsilon_{ij} {q_j\over q^2}\delta(q_z)~,$ where $i,j$ are in the transverse plane to the solenoid and $A_z$ vanishes.
This arises from the external current
$J_{{\rm ext}\ i}\sim  q_i\delta(q_z) $.

If we ignore the backreaction and study how the electrons react to the solenoid, we find to leading order in the magnetic field flux the induced  current $\expval{J_i}\sim K_{ij}A_j$.
We learn that there is no current at long distances, since the large $q$ limit is a polynomial in momentum space.
At short distances, though, we find
$\expval { J_i}\sim \epsilon_{ij}\grad_j \log|\bm r|$, with $\bm r$ in the transverse plane.
This is valid for $|\bm r|\ll k_F^{-1}$.
This leads to a log-divergent total current outside of the solenoid at short distances. This is precisely the source of our non-analyticity. In fact the current is finite in the limit of vanishing solenoid flux, signaling a breakdown in linear response theory. (Of course, being that it is a short distance divergence in the total current,  it is not universal and applies only in special circumstances, as we explain in the text.)

We now review the calculation that leads to~\eqref{eq:KTstatic}. Analogously to the electric calculation we get
\begin{align}
    \chi^{R}_{ij}(\bm q,0)
    =
    -\qty(\frac{e\hbar}{m})^2
    \int\frac{d^3k}{(2\pi)^3}
    \Big(k_i+\frac{q_i}{2}\Big)\Big(k_j+\frac{q_j}{2}\Big)
    \frac{\Theta(k_F-k)-\Theta(k_F-\abs{\bm k+\bm q})}{\xi_{\bm k}-\xi_{\bm k+\bm q}}.
    \label{eq:staticChiGeneral}
\end{align}
Since we already know the tensor structure, to simplify the calculation
we choose the momentum transfer along the $z$ axis
$\bm q=q\,\hat{\bm z}
$
and calculate $\chi^R_{xx}(q,0)$:
\begin{align}
     & \chi^{R}_{xx}(q,0)
    =
    -\qty(\frac{e\hbar}{m})^2
    \int\frac{d^3k}{(2\pi)^3}
    k_x^2\;
    \frac{\Theta(k_F-k)-\Theta(k_F-\abs{\bm k+\bm q})}{\xi_{\bm k}-\xi_{\bm k+\bm q}}
    \\ &=
    \frac{e^2}{m}\int\frac{d^3k}{(2\pi)^3}\;
    \frac{k_x^2}{q^2+2qk_z}\,
    \Big[\Theta(k_F-k)-\Theta(k_F-\abs{\bm k+\bm q})\Big].
    \label{eq:chi_xx_start}
\end{align}
The calculation is elementary in cylindrical coordinates and quickly reduces to
\begin{align}
    \chi^{R}_{xx}(q,0)
    =
    \frac{e^2}{m}\frac{1}{16\pi^2}
    \int_{-k_F}^{k_F}dk_z\;
    \frac{(k_F^2-k_z^2)^2}{q^2-4k_z^2}.
    \label{eq:chi_xx_simplified}
\end{align}
For $q<2k_F$ the denominator has poles at $k_z=\pm q/2$ inside the interval; the integral is understood as a Cauchy principal value, which is exactly what the retarded $\omega\to 0^+$ limit produces.
We find:
\begin{align}
    \chi^R_{xx}(q,0)
     & =
    \frac{ne^2}{m}\,\frac{3}{32}\frac{1}{z}
    \left[
        -2z^3+\frac{10}{3}z+(1-z^2)^2\ln\abs{\frac{1+z}{1-z}}
        \right].
\end{align}
The full transverse kernel is $K_T=K_{xx}=\chi^R_{xx}+ne^2/m$, so
\begin{align}
    \frac{K_T(q,0)}{ne^2/m}
     & =
    \frac{3}{8}(1+z^2)-\frac{3}{16z}(1-z^2)^2\ln\abs{\frac{1+z}{1-z}}
    \equiv f_T(z).
\end{align}
In summary, the full static kernel (including the seagull contact term) is purely transverse:
\begin{align}
    K_{ij}(\bm q,0)
    =
    \qty(\delta_{ij}-\frac{q_i q_j}{q^2})\frac{ne^2}{m}\,f_T\!\qty(\frac{q}{2k_F}),
    \qquad
    K_L(q,0)=0.
\end{align}

\section{Derivation of 2D current in the presence of AB flux}
\label{app:cur_AB_2d}
We want to compute the following quantity
\begin{gather}
    j^\theta(r) = \sum_{l \in \mathbb{Z} +\nu} l \int\limits^1_0 \frac{dk}{2\pi} k J^2_{|l|}(kr)  = \frac{1}{2\pi r^2} \sum_{l \in \mathbb{Z} +\nu} l  \int\limits^r_0 dt \, t J^2_{|l|}(t)
\end{gather}
Let us denote
\begin{gather}
    \mathcal{J}_\nu(r) = \sum_{l \geq 0} (l + \nu) J^2_{l + \nu}(r)
\end{gather}
Differentiating this expression termwise and use $2J_\alpha' = J_{\alpha-1}-J_{\alpha+1}$ and
$\frac{2\alpha}{r}J_\alpha=J_{\alpha-1}+J_{\alpha+1}$:
\[
    \mathcal{J}_\nu'(r)=2 \sum_{l\ge0}(l+\nu)\, J_{l+\nu}J'_{l+\nu}
    =\frac r2\sum_{l\ge0}\big(J_{l+\nu-1}^2-J_{l+\nu+1}^2\big)
    =\frac r2\big(J_{\nu-1}^2+J_\nu^2\big),
\]
Now integrating this expression we get
\[
    \mathcal{J}_\nu(r)=\Big(\nu+\frac{r^2}{2}\Big)J_\nu^2(r)+\frac{r^2}{2}J_{\nu+1}^2(r)
    -\frac{(2\nu+1)r}{2}J_\nu(r)J_{\nu+1}(r).
\]
Now plugging this expression in the equation for current and integrating over $r$ we arrive at
\begin{gather}
    j^\theta(r) =  \frac{1}{2\pi r^2}\left(S_\nu - S_{1 - \nu}\right), \notag\\
    S_{\nu}(r)= \frac{2r^{2}-2\nu^{2}+5\nu}{6}\,J_{\nu}^{2}(r)
    +\frac{2r^{2}-2\nu^{2}+3\nu-1}{6}\,J_{\nu+1}^{2}(r)\\
    +\left(\frac{\nu(2\nu-1)(\nu-1)}{3r}-\frac{(2\nu+1)r}{3}\right)
    J_{\nu}(r)J_{\nu+1}(r).
\end{gather}

\bibliographystyle{JHEP}
\bibliography{main}

\end{document}